# Development of a single-parameter spring–dashpot rolling friction model for coarse-grained DEM

Putri Mustika Widartiningsih[a,1], Yoshiharu Tsugeno[a,1], Toshiki Imatani[a],

Yuki Tsunazawa[b], Mikio Sakai[a*]

[a] Department of Nuclear Engineering and Management, The University of Tokyo, 7-3-1 Hongo, Bunkyo-ku, Tokyo 113-8656, Japan

[b] Geological Survey of Japan, National Institute of Advanced Industrial Science and Technology (AIST), Ibaraki 305-8567, Japan

\* Corresponding author:

Mikio Sakai

mikio_sakai@n.t.u-tokyo.ac.jp

TEL +81-3-5841-6977

FAX +81-3-5841-6981

[1] Equal contribution

**Abstract**

Simulating granular materials composed of non-spherical particles remains a major challenge in discrete element method (DEM) simulations due to the complexity of contact detection and rotational dynamics, rendering large-scale simulations computationally prohibitive. To address this limitation, rolling friction is commonly introduced as an approximation to account for particle shape effects by applying a resistive torque to spherical particles. Among existing rolling friction formulations, the spring–dashpot (S–D) type model is widely recognized for its numerical stability and realistic representation of rolling resistance. However, conventional S–D models require multiple empirical parameters that must be calibrated in an interdependent manner, leading to increased experimental effort, parameter ambiguity, and uncertainty in practical applications. To overcome these issues, this study proposes a new S–D type rolling friction model that reduces the parameter set to a single physically meaningful quantity: the critical rolling angle. Derived from theoretical considerations, this parameter characterizes the transition from static to rolling motion at particle contacts. The use of a single parameter simplifies implementation and eliminates the need for extensive calibration. Stability analysis demonstrates that the proposed model allows particles to reach a physically consistent equilibrium state without spurious rotational oscillations. For large-scale applications, the model is further integrated into a coarse-grained DEM framework. Validation using DEM–CFD simulations of an incinerator system confirms that the proposed approach successfully reproduces the macroscopic behavior of the original particle system. Overall, this study enhances the applicability of DEM for industrial-scale simulations involving non-spherical particles.

*1. Introduction*

The discrete element method (DEM), originally developed by Cundall and Strack [1], has grown into a powerful computational tool for analyzing granular flows in complex industrial systems. By resolving the fundamental forces and torques governing individual particle dynamics, DEM achieves both versatility and accuracy in predicting bulk particle behavior. Over the past decade, extensive review studies highlight its successful implementation across multiple disciplines, including powder transport [2], geotechnics [3], agriculture [4], nuclear engineering [5], and pharmaceuticals [6,7]. Modeling industrial particulate flow presents unique computational challenges, as they often require simultaneous modeling vast number of particles with diverse physical properties and interactions with geometrically complex boundaries [8–11]. These demanding applications necessitate both algorithmic efficiency and numerical robustness [12,13].

Conventionally, DEM utilizes spherical particles to model grains due to their efficiency and robustness of the associated algorithms [14]. However, this simplification poses a major limitation when applied to non-spheres, which are prevalent in most real-world applications [15–24]. Particle shape plays a critical role in determining system behavior [25–27]. Spherical particles generally exhibit simpler and predictable contact interactions, such as rolling and sliding. In contrast, non-spherical particles—with elongated bodies and flat features—tend to interlock more easily, forming shear-resistant structures. A deeper understanding of this shape effects can improve process efficiency, optimize equipment design, and enhance control strategies.

To represent real particle shapes, the DEM has been extended to incorporate non-spherical particles through several techniques including ellipsoids [28–32], super-quadrics [33–36], polyhedral [37–39], and multi-spheres [40–42]. While these advanced modeling approaches enable more physically accurate simulations, they introduce substantial computational overhead. Modeling non-spherical particles requires complex calculations for rotational motion and contact detection. These intensive computations make large-scale industrial simulations with realistic particle shapes computationally prohibitive.

To address this, researchers often implement a rolling friction approach to approximate non-spherical particle behavior [18,43–47]. This approach is motivated by the fact that non-spherical particles inherently resist rolling due to irregular shape features that act as physical barriers. In DEM, rolling friction is defined as a torque applied to each contacting particle pair that resists rolling motion, resembling the bulk behavior of non-spherical particles. This approach captures essential shape-dependent phenomena while preserving computational efficiency of spherical particles [43]. This makes rolling friction particularly valuable for large-scale simulations.

Several techniques for modeling rolling friction in DEM simulations have been proposed, such as the directional constant torque (DCT) model [48] and the spring–dashpot (S–D) type model [49]. DCT model has been widely adopted in DEM simulations. This model applies a constant opposing torque regardless of the contact dynamics. Its primary advantage lies in its computational simplicity, and it uses a single rolling friction coefficient to manage resistance. However, this feature can cause numerical instability once particle approaches stationary. Specifically, the constant torque may overestimate the resistance required to achieve equilibrium, potentially compromising accuracy [50]. The S–D type model addresses the stability issue by introducing a viscous damping torque. The torque magnitude is controlled by the relative angular displacement and the relative angular velocity, generating an adaptive torque mechanism. This means the total rolling friction torque changes based on the system's state, rather than being a constant value like in the DCT model. As a result, the S–D type model improves stability and provides a more realistic representation of particle behavior. However, its reliance on multiple artificial parameters complicates implementation. The original S–D model, proposed by Iwashita and Oda [49], requires four parameters, later modified by Luding to include three parameters for rolling friction and three for twisting friction [51]. Further, Jiang et al. have reduced the number of parameters through theoretical approaches, yet the model still requires at least two free parameters [52]. For practical applications, especially in large-scale simulations, a model with a single parameter is desired.

Rolling friction parameters are typically determined through indirect calibrations methods, which involve conducting laboratory experiments to measure specific bulk properties of the material [53–55]. The rolling friction parameters are iteratively adjusted until the simulated bulk response aligns with the experimental data. For reliable calibration, each experiment should be sensitive to only one parameter [56]. When additional parameters are introduced, complexity grows exponentially as each parameter requires independent characterization while also necessitating analyses of their collective influence on the system's behavior. Moreover, different parameter combinations can produce nearly identical responses, leading to ambiguity in calibration. Therefore, one parameter is highly preferable.

To address these challenges, this study proposes a modified S–D type model that uses a single parameter: the critical rolling angle. This parameter represents critical static-to-rolling transition condition and is derived from a theoretical framework that elucidates the rolling friction mechanism at the particle contact surface. By relying on one parameter, the proposed model eliminates the need for extensive calibration. Furthermore, the simplicity and robustness of the proposed model make it particularly suitable for large-scale systems, where the particle flow patterns in industrial scenarios are often complex. The use of a single tunable parameter allows for straightforward adjustment to diverse conditions.

For large-scale DEM simulations, a coarse-grained model [57] is frequently applied. The coarse-grained DEM reduces the computational load by particle grouping approach. This allows simulations to track representative particles instead of every individual grain, cutting computation time substantially. To preserve the overall particle dynamics, forces acting on the coarse-grained particles are appropriately scaled. Researchers have successfully applied coarse-grained DEM in numerous industrial scenarios, such as fluidized beds [58–61], granular sedimentation [62], die-filling [60,63], granular mixers, bead mills [64], and multiphase gas-solid-liquid flows [65]. Recent studies have demonstrated the adequacy of the combination of rolling friction model with coarse-grained DEM in various systems, such as dense medium cyclone [66], fluidized bed [67], granular heap test [68], and rotary kiln [69]. Despite these advancements, most studies rely on the DCT model, which is prone to instability issues. This emphasizes the need for further advancement in the integration of rolling friction with coarse-grained DEM. Therefore, the second stage of this study is to integrate the proposed rolling friction model with the coarse-grained DEM. This novel combination will enhance the applicability of coarse-grained DEM.

The primary objective of this study is to develop a DEM framework that enables large-scale simulations of non-spherical particles while maintaining computational efficiency. This aim is addressed in two stages: (i) development of a rolling friction model based on the S–D type requiring a single parameter, and (ii) integration of the proposed rolling friction model with the coarse-grained DEM. The proposed model's stability and scalability are evaluated across two scenarios: (i) granular heap tests to assess the stability of the proposed model in quasi-static conditions, and (ii) incinerator with a control plate to evaluate the applicability of the model in an industrial-relevant system. For the incinerator simulations, validation involves comparison of particle behavior between the original and coarse-grained models. A 10% deviation threshold between the original and coarse-grained models is implemented, reflecting consensus in DEM literature where deviations below this threshold have demonstrated acceptable representation of bulk particle behavior for industrial applications [9,70–72].

## 2. Numerical modeling

This study employs the flexible Eulerian–Lagrangian method with an implicit algorithm (FELMI) [13] to simulate gas–solid systems at an industrial scale. FELMI is a coupled DEM–CFD framework that integrates a coarse-grained discrete element method (DEM) for modeling large-scale solid particle dynamics with computational fluid dynamics (CFD) for resolving the fluid phase. In addition, signed distance functions (SDF) are used to represent particle–wall interactions, while the immersed boundary method (IBM) is employed to model fluid–wall interactions in complex geometries. In the present study, a rolling resistance model is newly incorporated into the FELMI framework to account for particle shape effects. The proposed rolling resistance formulation is applied to both the original and coarse-grained DEM, enabling stable and efficient simulations of granular systems composed of non-spherical particles.

### 2.1 Solid phase

#### 2.1.1 Modeling of translational motion

In FELMI, the solid phase is modeled using the DEM. Based on the Newton's second law of motion, the translational motion of solid particles is calculated as follows:

$$m\frac{d\boldsymbol{v}}{dt} = \sum \boldsymbol{F}_C + \boldsymbol{F}_D - V_p \nabla p + m\boldsymbol{g}, \tag{1}$$

where $m$, $\boldsymbol{v}$, $\boldsymbol{F}_C$, $\boldsymbol{F}_D$, $V_p$, $p$ and $\boldsymbol{g}$ are particle mass, particle velocity, contact force, solid–fluid interaction force, particle volume, pressure, and gravitational acceleration.

The contact force acting on a particle is decomposed into normal and tangential components:

$$\boldsymbol{F}_C = \boldsymbol{F}_{C_n} + \boldsymbol{F}_{C_t} \tag{2}$$

Then, $\boldsymbol{F}_{C_n}$ and $\boldsymbol{F}_{C_t}$ are expressed as

$$\boldsymbol{F}_{C_n} = -k_n \boldsymbol{\delta}_n - \eta_n \boldsymbol{v}_{r_n}, \tag{3}$$

$$\boldsymbol{F}_{C_t} = \begin{cases} -k_t \boldsymbol{\delta}_t - \eta_t \boldsymbol{v}_{r_t} & |\boldsymbol{F}_{C_t}| \leq \mu |\boldsymbol{F}_{C_n}| \\ -\mu |\boldsymbol{F}_{C_n}| \frac{\boldsymbol{v}_{r_t}}{|\boldsymbol{v}_{r_t}|} & |\boldsymbol{F}_{C_t}| > \mu |\boldsymbol{F}_{C_n}| \end{cases}, \tag{4}$$

where $k$, $\boldsymbol{\delta}$, $\eta$, $\boldsymbol{v}_r$ and $\mu$ indicate the spring constant, displacement, damping coefficient, relative velocity, and coefficient of friction, respectively. Subscripts $n$ and $t$ refer to the normal and tangential components.

The drag force acting on a particle is expressed as:

$$\boldsymbol{F}_D = \frac{\beta V_p}{1-\varepsilon} (\boldsymbol{u}_g - \boldsymbol{v}), \tag{5}$$

where $\beta$, $\varepsilon$ and $\boldsymbol{u}_g$ indicate the interphase momentum transfer coefficient, the void fraction, and the gas velocity, respectively. The void fraction is defined as the ratio of the void volume to the CFD grid volume, expressed as:

$$\varepsilon = 1 - \frac{\sum V_p}{V_{grid}}. \tag{6}$$

As well as the previous solid–fluid coupling simulations, the drag force is calculated by using the interphase momentum transfer coefficient, which depends on the particle concentration regimes within a CFD grid.

### 2.1.2. Modeling of rotational motion

For rotational motion, a rolling resistance moment is incorporated. The governing equation is given by Eq. (7).

$$I\frac{d\boldsymbol{\omega}}{dt} = \sum \boldsymbol{T} + \sum \boldsymbol{M}, \tag{7}$$

where $I$, $\boldsymbol{\omega}$, $\boldsymbol{T}$, and $\boldsymbol{M}$ denote the moment of inertia, angular velocity, external torque, and rolling resistance moment, respectively. The proposed rolling resistance follows a spring–dashpot formulation, consisting of an elastic resistance and a viscous damping term, similar in spirit to the models of Iwashita and Oda [49], and Jiang et al. [52, 73]. A yield criterion is introduced such that the rolling resistance moment is limited by a reference rolling moment $\boldsymbol{M}_{reff}$. When the computed moment exceeds $\boldsymbol{M}_{reff}$, the rolling resistance is capped to satisfy the threshold condition. Accordingly, the rolling resistance moment is expressed as:

$$M = \begin{cases} -k_\theta \boldsymbol{\theta}_{roll} - \eta_\theta \boldsymbol{\omega}_{roll} & if\ |M| \leq |M_{reff}| \\ M_{reff} & if\ |M| > |M_{reff}| \end{cases}, \tag{8}$$

where $k_\theta$ is the angular stiffness, $\boldsymbol{\theta}_{roll}$ is the relative rolling angular displacement of the two spheres, $\eta_\theta$ is the angular viscous damping, and $\boldsymbol{\omega}_{roll}$ is the relative rolling angular velocity, respectively.

**Fig. 1** schematically illustrates the rolling resistance and the distribution of the normal contact stress within the contact area. To evaluate the rolling interaction, the normal spring–dashpot system is conceptually decomposed into an infinite number of infinitesimal elements that are continuously distributed over the contact area. Assuming a circular contact area and a linearized stress distribution, the spring and dashpot coefficients per unit area are defined as:

$$K_n = \frac{k_n}{\pi R_c^2}, \tag{9}$$

$$H_n = \frac{\eta_n}{\pi R_c^2}, \tag{10}$$

where $K_n$, and $H_n$ denote the normal spring stiffness and viscous damping coefficient per unit contact area, respectively, and $R_c$ denotes the radius of the circular contact area. The stress distribution in the contact area is expressed as:

$$\begin{aligned}\boldsymbol{\sigma} &= \bar{\boldsymbol{\sigma}}_n + \boldsymbol{\sigma}_{roll}(z) \\ &= \bar{\boldsymbol{\sigma}}_n - K_n \boldsymbol{\theta}_{roll} z - H_n \boldsymbol{\omega}_{roll} z \end{aligned} \tag{11}$$

where $\bar{\boldsymbol{\sigma}}_n$ is the average normal stress, which given by:

$$\bar{\sigma}_n = \frac{F_{C_n}}{\pi R_c^2}. \tag{12}$$

The rolling resistance moment can be calculated by integrating the stress distribution across strips within the contact area.

$$\begin{aligned} \boldsymbol{M} &= \int_{-R_c}^{R_c} (\boldsymbol{\sigma}_{roll}(z)) z \left(2\sqrt{R_c^2 - z^2}\right) dz \\ &= 2\int_{-R_c}^{R_c} (-K_n \boldsymbol{\theta}_{roll} z - H_n \boldsymbol{\omega}_{roll} z) z \sqrt{R_c^2 - z^2} dz \\ &= -4(K_n \boldsymbol{\theta}_{roll} + H_n \boldsymbol{\omega}_{roll}) \int_0^{R_c} z^2 \sqrt{R_c^2 - z^2} dz \\ &= -4(K_n \boldsymbol{\theta}_{roll} + H_n \boldsymbol{\omega}_{roll}) \frac{\pi R_c^4}{16} \\ &= -\frac{R_c^2}{4}(k_n \boldsymbol{\theta}_{roll} + \eta_n \boldsymbol{\omega}_{roll}) \end{aligned} \tag{13}$$

Thus, the angular stiffness and angular viscous damping in the rolling direction are related to the spring constant and the damping coefficient in the contact force calculation.

$$k_\theta = \frac{R_c^2}{4} k_n, \tag{14}$$

$$\eta_\theta = \frac{R_c^2}{4} \eta_n. \tag{15}$$

Here, it is considered that rolling begins when the contact starts to detach, i.e., when there is a region where the stress is zero. The point where the stress firstly becomes zero is the location where $z = R_c$, as shown in Fig. 1.

$$\sigma_{min} = \bar{\sigma}_n + \sigma_{roll}(R_c)$$

$$= \frac{F_{C_n}}{\pi R_c^2} - K\boldsymbol{\theta}_{roll}R_c - H\boldsymbol{\omega}_{roll}R_c \tag{16}$$

$$= 0$$

Based on equations (13) and (16), the reference rolling resistance moment can be expressed as:

$$|\boldsymbol{M}_{ref}| = \frac{|\boldsymbol{F}_{C_n}|R_c}{4}. \tag{17}$$

As illustrated in Fig. 2, a particle placed on an inclined surface can remain stationary without rolling when the rolling resistance moment counterbalances the contact force from the slope and the gravitational torque acting on the particle. Under quasi-static conditions with a single contact between the particle and the inclined surface, the torque balance about the particle center provides a criterion for the onset of rolling. Accordingly, the threshold of the rolling resistance moment can be related to the inclination angle at which rolling initiates, as expressed by

$$|\boldsymbol{M}_{ref}| = |\boldsymbol{F}_{C_n}|r^*\tan\phi_0, \tag{18}$$

where $r^*$ denotes the effective particle radius, and $\phi_0$ denotes the critical slope angle at which rolling motion begins. Based on equations (17) and (18), the radius of the contact area is given by:

$$R_c = 4r^*\tan\phi_0. \tag{19}$$

As described above, the proposed rolling resistance model requires only the slope angle at which particles begin to roll as an additional parameter, alongside the spring constant and damping coefficient used to calculate the contact force. Since the slope angle can be obtained from simple experiments, the proposed model does not any arbitrary parameters. Furthermore, by introducing the threshold for the rolling resistance moment, stationary particles will not begin to rotate or move due to rotational force.

*2.1.3 Coarse-grained model with a rolling resistance model*

In FELMI, to enable simulation of large-scale solid particle systems, a coarse-grained DEM is employed. In the coarse-grained DEM, a group of original particles are assumed as one coarse grained particle. To represent a scale factor between a coarse-grained particle and their original particles, a coarse-grain ratio $l$ is introduced, where the diameter of a coarse-grained particle is $l$ times larger than that of an original particle to ensure mass conservation. Regarding the particle properties, the relationships between the coarse-grained and the original particle properties can be expressed as:

$$r_{CGM} = lr_O, \tag{20}$$

$$m_{CGM} = l^3 m_O, \tag{21}$$

where $r$ is the particle radius. The subscripts $CGM$ and $O$ stand for the coarse-grained particle and the original particle, respectively.

As well as our previous studies, the translational motion of a coarse-grained particle is assumed to be the average of its original particles. Regarding the translational motion, thus, the relationship between the original and coarse-grained particle properties is expressed as:

$$\bar{\boldsymbol{v}}_O = \boldsymbol{v}_{CGM}, \tag{22}$$

Based on these relationships, the governing equation of the translational motion of the coarse-grained particle is given by:

$$\begin{aligned} m_{CGM}\frac{d\boldsymbol{v}_{CGM}}{dt} &= \sum \boldsymbol{F}_{C_{CGM}} - V_{CGM}\nabla p + \boldsymbol{F}_{D_{CGM}} + m_{CGM}\boldsymbol{g} \\ &= l^3 \sum \bar{\boldsymbol{F}}_{C_O} - l^3 V_O \nabla p + l^3 \bar{\boldsymbol{F}}_{D_O} + l^3 m_O \boldsymbol{g}. \end{aligned} \tag{23}$$

As illustrated in **Fig. 2**, the rotational motion of a coarse-grained particle is assumed to follow the average angular velocity of its constituent particles, based on the assumption that the original particles within the coarse-grained particle rotate around their centers of mass with equal angular velocity. The following relationships are defined for angular kinetic energy in both the coarse-grained particle and the original particles.

$$\frac{1}{2}I_{CGM}\boldsymbol{\omega}_{CGM}^2 = l^3 \frac{1}{2}I_O \bar{\boldsymbol{\omega}}_O^2, \tag{24}$$

$$\begin{aligned} I_{CGM} &= \frac{2}{5}m_{CGM}r_{CGM}^2 \\ &= \frac{2}{5}(l^3 m_O)(lr_O)^2 \\ &= l^5 \frac{2}{5}m_O r_O^2 \\ &= l^5 I_O \end{aligned} \tag{25}$$

From Eqs. (24) and (25), the following relationship can be derived.

$$\omega_{CGM} = \frac{1}{l}\overline{\omega}_O. \tag{26}$$

As for strain energy in the rolling, the following relationships are defined for the coarse-grained particle and the original particles.

$$\frac{1}{2}k_{\theta_{CGM}}\boldsymbol{\theta}_{roll_{CGM}}^2 = l^3 \frac{1}{2}k_{\theta_O}\overline{\boldsymbol{\theta}}_{roll_O}^2. \tag{27}$$

Based on Eqs. (16) and (21), the angular stiffness for the coarse-grained particle can be transformed as follows. The slope angle at which the coarse-grained particle begins to roll is assumed to be the same as the slope angle at which original particles begin to roll.

$$\begin{aligned}k_{\theta_{CGM}} &= \frac{R_{c_{CGM}}^2}{4}k_{CGM} \\ &= \frac{(4r_{CGM}^*\tan\phi_0)^2}{4}k_{CGM} \\ &= \frac{(4lr_O^*\tan\phi_0)^2}{4}l^3 k_O \\ &= 4l^5 r_O^* \tan^2\phi_0 k_O.\end{aligned} \tag{28}$$

In contrast, the angular stiffness for the original particle can be transformed as follows.

$$\begin{aligned}k_{\theta_O} &= \frac{R_{c_O}^2}{4}k_O \\ &= \frac{(4r_O^*\tan\phi_0^O)^2}{4}k_O \\ &= 4r_O^*\tan^2\phi_0 k_O.\end{aligned} \tag{29}$$

Finally, the following relationship for the angular stiffness can be obtained.

$$k_{\theta_{CGM}} = l^5 k_{\theta_O}. \tag{30}$$

Accordingly, based on Eqs. (27) and (30), the following relationship for the relative angular displacement can be obtained.

$$\boldsymbol{\theta}_{roll\,CGM} = \frac{1}{l}\overline{\boldsymbol{\theta}}_{roll\,O}. \tag{31}$$

In the same manner with the angular stiffness, based on Eqs. (15) and (19), the angular viscous damping for the coarse-grained particle can be transformed as follows.

$$\begin{aligned}\eta_{\theta_{CGM}} &= \frac{R_{c_{CGM}}^2}{4}\eta_{CGM} \\ &= \frac{(4r_{CGM}^*\tan\phi_0)^2}{4}\eta_{CGM} \\ &= \frac{(4lr_O^*\tan\phi_0)^2}{4}l^3\eta_O \\ &= 4l^5 r_O^{*}\tan^2\phi_0\eta_O. \end{aligned} \tag{32}$$

In contrast, the angular viscous damping for the original particle can be transformed as follows.

$$\begin{aligned}\eta_{\theta_O} &= \frac{R_{c_O}^2}{4}\eta_O \\ &= \frac{(4r_O^*\tan\phi_0^O)^2}{4}\eta_O \\ &= 4r_O^{*}\tan^2\phi_0\eta_O. \end{aligned} \tag{33}$$

Finally, the following relationship for the angular stiffness can be obtained.

$$\eta_{\theta_{CGM}} = l^5\eta_{\theta_O}. \tag{34}$$

Based on above relationships, the rolling resistance moment of the coarse-grained particle

is given by:

$$\begin{aligned}
\boldsymbol{M}_{CGM} &= -k_{\theta_{CGM}}\boldsymbol{\theta}_{roll_{CGM}} - \eta_{\theta_{CGM}}\boldsymbol{\omega}_{roll_{CGM}}, \\
&= -l^5 k_{\theta_O}\frac{1}{l}\overline{\boldsymbol{\theta}}_{roll_O} - l^5 \eta_{\theta_O}\frac{1}{l}\overline{\boldsymbol{\omega}}_{roll_O} \\
&= l^4\left(-k_{\theta_O}\overline{\boldsymbol{\theta}}_{roll_O} - \eta_{\theta_O}\overline{\boldsymbol{\omega}}_{roll_O}\right) \\
&= l^4 \overline{\boldsymbol{M}}_O.
\end{aligned} \quad (35)$$

Hence, the governing equation of the rotational motion of the coarse-grained particle is given by:

$$\begin{aligned}
I_{CGM}\frac{d\boldsymbol{\omega}_{CGM}}{dt} &= \sum \boldsymbol{T}_{CGM} + \sum \boldsymbol{M}_{CGM} \\
&= l^4 \sum \overline{\boldsymbol{T}}_O + l^4 \sum \overline{\boldsymbol{M}}_O.
\end{aligned} \quad (36)$$

*2.2 Fluid phase*

In FELMI, the fluid phase is calculated based on the CFD. In the same manner as the conventional DEM-CFD, based on the local-average technique [75], the fluid continuity and Navier-Stokes equations for an incompressible fluid are given by:

$$\frac{\partial \varepsilon}{\partial t} + \nabla \cdot (\varepsilon \boldsymbol{u}_g) = 0, \quad (37)$$

$$\frac{\partial(\varepsilon \rho_g \boldsymbol{u}_g)}{\partial t} + \nabla \cdot (\varepsilon \rho_g \boldsymbol{u}_g \boldsymbol{u}_g) = -\varepsilon \nabla p - \boldsymbol{f} + \nabla \cdot (\varepsilon \boldsymbol{\tau}_g) + \varepsilon \rho_g \boldsymbol{g}, \quad (38)$$

where $\rho_g$, $\boldsymbol{f}$, and $\boldsymbol{\tau}_g$ are the gas density, drag force acting on the fluid, and the viscous tensor. The interaction between solid and fluid phases is evaluated based on the Newton's third law of motion. That is, the drag force acting on the fluid is calculated by:

$$f = \frac{\sum F_D}{V_{grid}}, \qquad (39)$$

where $V_{grid}$ is a volume of a fluid grid. The drag force is determined using the interphase momentum transfer coefficient, depending on the particle concentration regimes within a fluid grid. In this study, the boundary between the dense regime and the dilute regime is determined by a void fraction of 0.8. The Ergun's model [76] is employed for the dense regime ($\varepsilon \leq 0.8$), while the Wen-Yu's model [77] is employed for the dilute regime ($\varepsilon > 0.8$).

$$\beta = \begin{cases} 150\dfrac{\eta_g(1-\varepsilon)^2}{\varepsilon d_o^2} + \dfrac{1.75\rho_g(1-\varepsilon)}{d_o}|v_{CGM} - u_g| & (\varepsilon \leq 0.8) \\ \dfrac{3}{4}C_D\dfrac{\varepsilon(1-\varepsilon)\rho_g}{d_o}|v_{CGM} - u_g|\varepsilon^{-2.65} & (\varepsilon > 0.8) \end{cases}, \qquad (40)$$

where $\eta_g$ is the gas viscosity, $d$ is the particle diameter, and $C_D$ is the drag coefficient, determined by:

$$C_d = \begin{cases} \dfrac{24}{Re_p}(1 + 0.15Re_p^{0.687}) & Re_p \leq 1000 \\ 0.44 & Re_p > 1000 \end{cases}, \qquad (41)$$

where the particle Reynolds number $Re_p$ is given as:

$$Re_p = \frac{\varepsilon\rho_g d_o}{\eta_g}|v_{CGM} - u_g| \qquad (42)$$

***2.3 Wall boundary***

In FELMI, the wall boundaries for solid and gas phases are modeled by the signed distance functions (SDF) and the immersed boundary method (IBM). The combination of the SDF and the IBM allows efficient modeling of arbitrarily complex-shaped or moving wall boundaries with a simple algorithm. Details and the applicability of the SDF and the IBM models in the DEM-CFD simulation has been demonstrated [78]. Here, we provide a brief description of the SDF and the IBM models.

In the SDF model, the wall boundary is expressed by a scalar field which is defined as

$$\phi(x) = s(x)d(x), \tag{43}$$

where, $s(x)$ is the sign, takes the values 1 and -1 for points inside and outside the calculation domain, respectively, and $d(x)$ is the minimal distance between a point and a wall boundary. The zero value of the SDF indicates the surface of the wall boundary. In the same manner as the calculation of particle–particle contact force, the particle–wall contact force is decomposed into normal and tangential components:

$$\boldsymbol{F}_{C_n}^{SDF} = l^3\left(-k_n \boldsymbol{\delta}_n^{SDF} |\nabla \phi| - \eta_n \boldsymbol{v}_{r_n}\right), \tag{44}$$

$$\boldsymbol{F}_{C_t}^{SDF} = \begin{cases} -k_t \boldsymbol{\delta}_t^{SDF} - \eta_t \boldsymbol{v}_{r_t} & \left|\boldsymbol{F}_{C_t}^{SDF}\right| \le \mu \left|\boldsymbol{F}_{C_n}^{SDF}\right| \\ -\mu \left|\boldsymbol{F}_{C_n}^{SDF}\right| \dfrac{\boldsymbol{v}_{r_t}}{|\boldsymbol{v}_{r_t}|} & \left|\boldsymbol{F}_{C_t}^{SDF}\right| > \mu \left|\boldsymbol{F}_{C_n}^{SDF}\right| \end{cases}. \tag{45}$$

In the normal component, the overlap between a particle and a wall boundary $\boldsymbol{\delta}_n^{SDF}$ is expressed as:

$$\boldsymbol{\delta}_n^{SDF} = \left(\phi - \frac{ld_0}{2}\right)\frac{\nabla \phi}{|\nabla \phi|}. \tag{46}$$

In the IBM, the fluid-wall interaction is evaluated by the volume-weighted average velocity:

$$\boldsymbol{u} = (1-\alpha)\boldsymbol{u}_g + \alpha \boldsymbol{U}_B, \tag{47}$$

where $\boldsymbol{U}_B$ is velocity of the wall and $\alpha$ is the local volume fraction occupied by the solid wall. When the IBM is considered in the calculation of the Navier-Stokes equation, an external forcing on fluid $\boldsymbol{f}_{IB}$ should be introduced in the Navier-Stokes equation. Thus, Eq. (38) is modified as:

$$\frac{\partial(\varepsilon \rho_g \boldsymbol{u}_g)}{\partial t} + \nabla \cdot (\varepsilon \rho_g \boldsymbol{u}_g \boldsymbol{u}_g) = -\varepsilon \nabla p - \boldsymbol{f} + \nabla \cdot (\varepsilon \boldsymbol{\tau}_g) + \varepsilon \rho_g \boldsymbol{g} + \boldsymbol{f}_{IB}, \tag{48}$$

with

$$\boldsymbol{f}_{IB} = \frac{\varepsilon \rho_g \alpha (\boldsymbol{U}_B - \boldsymbol{u}^*)}{\Delta t}, \tag{49}$$

where $\boldsymbol{u}^*$ is the temporal velocity field before being modified by the IBM.

## *2.4 Calculation stability*

To ensure the numerical stability, it is crucial to determine appropriate time step. In the DEM employing the Voigt model, the time step for calculating contact force is often determined based on the oscillation period of the spring–mass system [79]. Specifically, the criteria for the time step are given by:

$$\Delta t \leq \frac{2\pi}{\Omega} \sqrt{\frac{m}{k}}, \tag{50}$$

where $\Omega$ is an empirical constant ranging from 5 to 20. Since the proposed rolling resistance model is based on the spring–dashpot system, the time step for calculating the rolling resistance should be determined in the same manner as for that for calculating contact force. Namely, by using the inertia and spring constant, the criteria for the time step derived from the rolling resistance model are expressed as:

$$\Delta t \leq \frac{2\pi}{\Omega}\sqrt{\frac{I}{k_\theta}}, \tag{51}$$

Therefore, to ensure the numerical stability of the DEM employing the Voigt model and the rolling resistance model proposed in this study, the time step should satisfy the following criteria.

$$\Delta t \leq min\left(\frac{2\pi}{\Omega}\sqrt{\frac{m}{k_n}}, \frac{2\pi}{\Omega}\sqrt{\frac{I}{k_\theta}}\right), \tag{52}$$

## 3. Stability test of the proposed rolling resistance model

Four simulation cases were conducted to assess the stability and adequacy of the proposed rolling resistance model. The model performance was evaluated through comparative simulations with an existing rolling resistance model, focusing on the collapse behavior of a particle assembly on a flat surface. To simplify the analysis and clarify the effect of rolling resistance, the collapse process was treated as a quasi-two-dimensional system.

### 3.1 Existing rolling resistance model

Several rolling resistance models have been proposed in the past studies. Among them, the directional constant torque (DCT) model [48] is widely used because of its simplicity. In the DCT model, a rolling resistance moment with a constant magnitude is applied in the direction opposite to the relative rotational motion at a contact. Consequently, the rolling resistance moment always opposes the relative rotation during particle–particle and particle–wall contacts. The rolling resistance moment is typically expressed as:

$$\boldsymbol{M}_{roll} = -\frac{\boldsymbol{\omega}_i - \boldsymbol{\omega}_j}{|\boldsymbol{\omega}_i - \boldsymbol{\omega}_j|}\mu_r r^* |\boldsymbol{F}_{C_n}|, \tag{55}$$

where $\mu_r$ is the coefficient of rolling friction. For brevity, this model is referred to as the existing model hereafter.

*3.2 Calculation system and conditions*

Following the verification system adopted in a previous study [50], an assembly of 100 particles was regularly arranged on a flat surface. Upon release, the particle assembly collapsed onto the surface under the action of gravity. **Fig. 4** illustrates the initial particle configuration, and the physical properties of the particles are summarized in **Table 1**. The particles were assumed to be monodispersed, with a diameter of 0.1 mm and a density of 2,500 kg/m$^3$, corresponding to typical glass beads. A soft-spring model was employed in the DEM simulations; therefore, the spring constant was set to 100 N/m. The coefficients of restitution and friction were set to 0.9 and 0.3, respectively, which are commonly used values in previous DEM studies. All simulations employed identical physical parameters, except for the critical rolling angle in the proposed rolling resistance model and the rolling friction coefficient in the existing rolling resistance model. The proposed rolling resistance model was applied in Cases 1–1 and 1–3, whereas the existing rolling resistance model was applied in Cases 1–2 and 1–4. The critical rolling angle was set to 0.10 rad and 0.20 rad in Cases 1-1 and 1-3, respectively, while the rolling friction coefficient was set to 0.05 and 0.10 in Cases 1-2 and 1-4, respectively. In all cases, the time step was set at $1.0 \times 10^{-6}$ s, and the total calculation time was 8.0 s to ensure the particle assemblies reached a steady state.

*3.3 Results and discussion*

First, under conditions of relatively low rolling resistance, the particle position distributions were visually compared. **Fig. 5** shows representative snapshots of the particle configurations at $t = 5.0$ s and $t = 8.0$ s for Cases 1–1 and 1–2. As no noticeable difference in the particle positions were observed between these two-time instants, it can be concluded that particle translational and rotational motions had largely ceased after $t = 5.0$ s in both cases.

To evaluate the stability of each system, the absolute angular velocity of particles was examined. **Fig. 6** shows color-coded snapshots of the angular velocity magnitude, focusing on regions where particles predominantly accumulate. In the proposed model (Case 1–1), little difference was observed in the distribution of the absolute angular velocity between $t = 5.0$ s and $t = 8.0$ s, indicating that the particle assembly had nearly reached a steady state, with both translational and rotational motions effectively suppressed. This behavior can be attributed to the proposed rolling resistance model, in which the rolling resistance moment progressively reduces particle rotation as a function of both the relative rolling displacement and the angular velocity. In contrast, in the existing model (Case 1–2), although the particle positions remained almost unchanged between $t = 5.0$ s and $t = 8.0$ s, a large number of particles retained finite angular velocities. This is because, in the existing model, the rolling resistance moment is determined solely by the contact force and always acts in the direction opposite to the particle rotation, which results in a gradual but persistent rotational motion rather than complete rotational stabilization.

For a quantitative assessment of the stability of each rolling resistance model, the transient change in the mean absolute angular velocity of particles in Cases 1–1 and 1–2 was compared, as shown in **Fig. 7**. In the proposed model (Case 1–1), the mean absolute angular velocity remained nearly zero throughout the observed period, whereas in the existing model (Case 1–2), relatively large fluctuations were observed. As shown in Fig. 5, the negligible difference in particle position distributions between $t = 5.0$ s and $t = 8.0$ s indicates that the observed fluctuations in the mean absolute angular velocity in Case 1–2 did not originate from particle rearrangements or translational motion. Instead, they can be attributed to the inherent characteristics of the existing rolling resistance model. Specifically, in the existing model, the rolling resistance moment is formulated as a constant torque proportional to the contact force and always acts opposite to the direction of rotation, which induces oscillatory rotational behavior in particles even under quasi-static conditions.

Next, under conditions of relatively high rolling resistance, the particle position distributions were also visually compared for Cases 1–3 and 1–4, as shown in **Fig. 8**. As in the low rolling resistance cases, little difference was observed in the particle position distributions between $t = 5.0$ s and $t = 8.0$ s, indicating that particle translational and rotational motions had largely ceased after $t = 5.0$ s in both cases.

**Fig. 9** shows color-coded snapshots of the absolute angular velocity magnitude in Cases 1–3 and 1–4, analogous to those presented in Fig. 6. In the proposed model (Case 1–3), the angular velocity distributions at $t = 5.0$ s and $t = 8.0$ s were nearly identical, indicating a stable rotational state. In contrast, in the existing model (Case 1–4), several particles exhibited noticeable differences in their absolute angular velocities between two time instants, despite the particle positions remaining nearly unchanged.

**Fig. 10** presents the transient change in the mean absolute angular velocity of particles in Cases 1–3 and 1–4, providing a quantitative assessment of the stability of each rolling resistance model. In the proposed model (Case 1–3), the mean absolute angular velocity decayed rapidly during the initial transient stage and remained nearly zero once the system settled into a quasi-static configuration, whereas the existing model (Case 1–4), persistent fluctuations in the mean absolute angular velocity were observed. Similar to the low rolling resistance conditions, this difference arises because the proposed model suppresses particle rotation through a rolling resistance moment that depends on both the relative rolling displacement and the angular velocity. In contrast, although the existing model also reduces particle rotation, oscillatory rolling motion persists due to the rolling resistance moment being modeled as a constant torque proportional to the contact force. Consequently, the proposed rolling resistance model exhibits superior numerical stability compared with the existing model, even under conditions of relatively high rolling resistance.

## *4. Validation test in an incinerator with a control plate*

### *4.1 Validation system and conditions*

To demonstrate the adequacy of the proposed rolling resistance model, validation tests were conducted using an incinerator equipped with a control plate. Macroscopic characteristics—including pressure drop, transient changes in total kinetic energy, the particle fraction above the control plate, and the angle of repose of the particle assembly on the control plate—were evaluated and compared between the original particle system and the coarse-grained particle system. **Fig. 11** shows a schematic of the incinerator with the control plate used in validation tests. The height, width, and depth of the calculation domain were 200 mm, 40 mm, and 40 mm, respectively. A control plate with a diameter of 23 mm was installed at a height of 140 mm from the bottom of the incinerator. The inlet gas velocity was 1.5 m/s for all cases. The top and bottom boundaries were defined as outlet and inlet boundaries, respectively. For fluid dynamics calculations, the grid size was chosen to be sufficiently larger than the particle diameter. Accordingly, the calculation domain was discretized into uniform grids with a grid size of 800 μm, resulting in 50, 250, and 50 grid cells in the x-, y-, z-directions, respectively.

The physical properties of the gas and solid particles used in the simulations are listed in **Table 2**. The gas phase was modeled as air, with a density of 1.0 kg/m$^3$ and a viscosity of $1.8 \times 10^{-5}$ Pa·s. The solid particles were assumed to be mono-dispersed and had a density of 2,500 kg/m³, representing typical glass beads. The spring constant, coefficient of friction, and coefficient of restitution were set to 100 N/m, 0.3, and 0.9, respectively. These physical parameters were kept identical for all simulation cases.

**Table 3** summarizes the calculation conditions. The simulation cases were classified into two groups: the coarse-grained DEM-CFD simulations without the proposed rolling resistance model (Case 2) and those incorporating the proposed rolling resistance model (Case 3). For each group, four simulation cases were conducted to evaluate the adequacy of the coarse-grained model, with and without the proposed rolling resistance model. Case 2–1 corresponds to the original particle system, with a particle diameter of 100 μm and a total of 2,560,000 particles. Cases 2–2 and 2–3 employ the coarse-grained model without the proposed rolling resistance model, with the coarse-grain ratios of 2.0 and 4.0, respectively. In these cases, the numbers of particles are set to 320,000 for Case 2–2 and 40,000 for Case 2–3, while maintaining the total particle mass equivalent to that of the original particle system (Case 2–1). Case 2–4 represents a system consisting of simply enlarged particles, without applying the coarse-grained model or the proposed rolling resistance model. In this case, both the particle diameter and the number of particles are identical to those in Case 2–3. Similarly, four calculation conditions were conducted for Case 3. Case 3–1 represents the original particle system, identical to Case 2–1. In Cases 3–2 and 3–3, the coarse-grained model with the proposed rolling resistance model is applied, using the coarse-grain ratios of 2.0 and 4.0, respectively. Case 3–4 represents a system with simply larger particles, without using the coarse-grained model but incorporating the proposed rolling resistance model. The time increments for the gas and particle phases were set to $5.0\times10^{-6}$ s and $1.0\times10^{-6}$ s, respectively, resulting in five particle sub-iterations per gas phase time step. As shown in Fig. 11, particles were initially placed randomly inside the incinerator, after which the gas flow was introduced from the bottom boundary.

### *4.2 Results and discussion*

#### *4.2.1 Case 2: coarse-grained DEM without rolling resistance model*

The macroscopic particle behavior in the incinerator was numerically investigated. **Fig. 12** shows representative snapshots of the particle behavior for Case 2. As shown in Fig. 12(a), referred to as the early stage, immediately after the gas flow is initiated, particles are lifted and dispersed throughout the incinerator. Subsequently, in the later stage, particles are primarily concentrated in the region above the control plate, as shown in Fig. 12(b). The particle behavior observed in the coarse-grained particle systems (Cases 2–2 and 2–3) is in good agreement with that of the original particle system (Case 2–1). In contrast, the particle behavior in the simply enlarged particle system without the coarse-grained model differs significantly from that of the original particle system: particles remain near the bottom region of the incinerator, and only a few reach the region above the control plate. This discrepancy is attributed to the difference in the fluid drag force acting on the particles, namely, fluid–particle interactions. Owing to this substantial deviation in particle behavior, subsequent quantitative comparisons are restricted to the original particle systems (Case 2–1) and the coarse-grained particle systems (Cases 2–2 and 2–3).

**Fig. 13** shows representative velocity distributions of particles and gas in the later stage, during which particles are primarily concentrated in the region above the control plate. The gas velocity is shown on the central cross-section, whereas the particles are visualized over the depth direction rather than on the central cross-section. As shown in Fig. 13(a), most particles are observed moving above the control plate, while some particles with little or no velocity accumulate on the control plate. The spatial distribution of particle velocity is qualitatively consistent between the original particle system (Case 2–1) and the coarse-grained particle systems (Cases 2–2 and 2–3). In addition, the gas velocity distribution in the original particle system is qualitatively similar to that in the coarse-grained particle systems. The control plate disrupts the upward gas flow, generating vortical structures and dispersing the gas velocity field. This flow disruption contributes to particle accumulation on the control plate. Overall, these results indicate that the coarse-grained model is capable of qualitatively reproducing the particle and gas flow behaviors observed in the original particle system within the incinerator.

To quantitatively assess the macroscopic dynamics of the particle system, the transient change in the total kinetic energy of particles in the incinerator is compared between the original particle system and the coarse-grained particle systems, as shown in **Fig. 14**. In both the original particle system (Case 2–1) and the coarse-grained particle systems (Cases 2–2 and 2–3), a pronounced increase in kinetic energy is observed immediately after the gas flow is initiated. Following this initial sharp increase, the total kinetic energy gradually increased. The temporal evolution of the total kinetic energy in the coarse-grained particle systems (Cases 2–2 and 2–3) shows good quantitative agreement with that in the original particle system (Case 2–1). For further quantitative comparison, the transient change in the particle fraction in the region above the control plate is examined, as shown in **Fig .15**. In both the original particle system (Case 2–1) and the coarse-grained particle systems (Cases 2–2 and 2–3), the particle fraction above the control plate gradually increases and then remains nearly constant. Overall, these results quantitatively demonstrate the adequacy of the coarse-grained model in reproducing both the transient behavior of the total kinetic energy of particles and the temporal evolution of the particle fraction above the control plate in the incinerator.

For a qualitative assessment of fluid dynamics behavior, **Fig. 16** compares the pressure drop between the original particle system (Case 2–1) and the coarse-grained particle systems (Cases 2–2 and 2–3). Although temporal fluctuations were observed in all cases, the pressure drop showed good agreement between the original particle system and the coarse-grained particle systems. This indicates that the coarse-grained particle systems exhibit fluidization behavior and macroscopic gas-solid interaction characteristics similar to those of the original particle system. The qualitative consistency in pressure drop suggests that the coarse-grained model can adequately reproduces the fluidization dynamics of the original particle system.

For a further comparison of particle dynamics, **Fig. 17** shows instantaneous snapshots of particles on the control plate at $t = 10.0$ s, corresponding to a quasi-steady state. In all cases, particles are observed to accumulate on the control plate. Qualitatively, the overall accumulation patterns appear similar among the three cases. To quantitatively assess the particle deposition behavior on the control plate, the angle of repose was determined by extracting the surface profiles of the particle pile in polar coordinates. A linear fit was applied to the radial height profile over an automatically selected interval that maximized the coefficient of determination, and the angle of repose was calculated from the fitted slope and statistically averaged over all directions. As summarized in Table 4, the angle of repose obtained using the coarse-grained model (Cases 2–2 and 2–3) was slightly smaller than that of the original particle (Case 2–1). This discrepancy indicates that simulations without a rolling resistance model are not sufficient to accurately reproduce particle deposition behavior on the control plate. These results highlight the necessity of incorporating the proposed rolling resistance model for capturing both the qualitative and quantitative characteristics of particle accumulation.

*4.2.2 Case 3: coarse-grained DEM with rolling resistance model*

This section focuses on validating the coarse-grained DEM incorporating the proposed rolling resistance model. **Fig. 18** shows representative snapshots of particle behavior for Case 3. Immediately after the gas flow was initiated, particles were lifted and dispersed throughout the incinerator. Subsequently, the particles predominantly accumulated in the region above the control plate. The macroscopic particle behavior in the coarse-grained particle systems (Cases 3–2 and 3–3) qualitatively agreed with that of the original particle system (Case 3–1). In contrast, the particle behavior in the system consisting of simply enlarged particles without the coarse-grained model (Case 3–4) differed significantly from that of the original particle system, with most particles remaining near the lower region of the incinerator. Consequently, as in Case 2, subsequent comparisons are restricted to the original particle system (Case 3–1) and the coarse-grained particles systems (Cases 3–2 and 3–3).

**Fig. 19** presents the velocity distribution of particles and gas in the region above the control plate. The gas velocity is shown on the central cross-section, whereas the particles are visualized over the depth direction rather than on the central cross-section. As shown in Fig. 19(a), most particles were observed flying above the control plate, whereas some particles with little or no velocity accumulated on the control plate. In comparison with Fig. 13, the number of particles accumulating on the control plate in Case 3 was noticeably higher than in Case 2. This difference can be attributed to the influence of the rolling resistance model. Specifically, when particles are in contact, the rolling resistance moment suppresses relative rolling motion between particles, thereby promoting particle accumulation on the control plate. The spatial distribution of particle velocity and position, including those of the accumulated particles, showed good qualitatively agreement between the original particle system (Case 3–1) and coarse-grained particle systems (Cases 3–2 and 3–3). In addition, the gas velocity distribution in the coarse-grained particle systems qualitatively reproduced that of the original particle system. Overall, these results indicate that the coarse-grained model incorporating rolling resistance can qualitatively reproduce the particle behavior observed in the original particle system within the incinerator.

A quantitative comparison was performed for the transient change in the total kinetic energy of particles in the incinerator and the particle fraction in the region above the control plate. As shown in **Fig. 20**, a sharp increase in the total kinetic energy was observed immediately after the onset of gas flow, followed by a gradual increase toward a quasi-steady state. The transient profiles of the total kinetic energy in the coarse-grained particle systems (Cases 3–2 and 3–3) closely agreed with those observed in the original particle system (Case 3–1). **Fig. 21** shows the transient change in the particle fraction in the region above the control plate. The evolution of the particle fraction in the coarse-grained particle systems (Cases 3–2 and 3–3) quantitatively matched that of the original particle system (Case 3–1). Overall, these results quantitatively demonstrate the adequacy of the coarse-grained DEM incorporating the rolling resistance model in reproducing both the transient change in the total kinetic energy of particles and the particle fraction above the control plate in the incinerator.

**Fig. 22** compares the pressure drop between the original particle system (Case 3–1) and the coarse-grained particle systems (Cases 3–2 and 3–3). Although temporal fluctuations were observed, the pressure drops in the coarse-grained particle systems (Cases 3–2 and 3–3) showed good qualitative agreement with that of the original particle system (Case 3–1). As in Case 2, this result indicates that the original and the coarse-grained particle systems exhibit similar fluidization behavior and macroscopic dynamic characteristics. The qualitative consistency of the pressure drop across the cases suggests that the coarse-grained model incorporating the rolling resistance model can adequately reproduce the fluidization dynamics and gas-solid interactions of the original particle system.

For a further comparison of particle deposition behavior, **Fig. 23** shows instantaneous snapshots of particles on the control plate at $t = 10.0$ s, corresponding to a quasi-steady state. As in Case 2, particles are observed to accumulate on the control plate, and qualitatively similar accumulation patterns are observed among the three cases. To quantitatively assess particle deposition on the control plate, the angle of repose was evaluated. As summarized in Table 4, the angle of repose obtained using the coarse-grained particle systems (Cases 3–2 and 3–3) were in close agreement with that of the original particle system (Case 3–1). This close consistency indicates that simulations incorporating the proposed rolling resistance model can accurately reproduce particle deposition behavior on the control plate. In contrast to the results obtained without the rolling resistance model (Case 2), these findings demonstrate that the proposed rolling resistance model is essential for capturing both the qualitative and quantitative characteristics of particle accumulation in the coarse-grained DEM simulations.

## 5. Conclusion

To address the challenge that industrial particles are generally non-spherical, extensive numerical studies have been conducted on modeling non-spherical particle behavior. Among the proposed approaches, rolling resistance models—which represent particle shape effects through rolling friction—have been recognized as a promising and computationally efficient method. In this study, a novel rolling resistance model was developed that requires only a single additional parameter obtained from simple experiments. The proposed model employs a spring–dashpot system to describe rolling motion kinetics, thereby improving numerical stability. By considering the relationship between rolling resistance and the normal contact stress distribution within the contact area, the spring resistance, viscous damping in rotation, and the maximum rolling resistance moment were consistently related to the contact force parameters used in DEM calculations. The stability of the proposed rolling resistance model was first examined by comparison with an existing rolling resistance model through simulations of particle assembly collapse on a flat surface. Analysis of particle position distributions and transient changes in angular velocity demonstrated that the proposed model exhibits superior rotational stability. Furthermore, the proposed rolling resistance model was incorporated into a coarse-grained DEM-CFD framework. Validation tests conducted in an incinerator with a control plat showed that the coarse-grained DEM-CFD simulations incorporating the proposed model successfully reproduced the macroscopic particle and gas behavior of the original particle system. Good agreement was obtained in terms of particle motion, transient changes in total kinetic energy, pressure drop, particle accumulation behavior, and the angle of repose on the control plate. These results confirm that the proposed rolling resistance model enables accurate and stable coarse-grained DEM-CFD simulations of gas-solid flows involving non-spherical particle effects. Overall, this study demonstrates that the proposed model provides a practical and reliable approach for large-scale industrial simulations, such as incinerator systems with internal structures.

***Acknowledgements***

The authors gratefully acknowledge financial support provided by JSPS KAKENHI (Grant No. 21H04870 and 21K19760), and JST SPRING (Grant No. JPMJSP2108).

**Fig. 1 Schematic diagram of the rolling resistance model and the distribution of the normal contact stress within the contact area.**

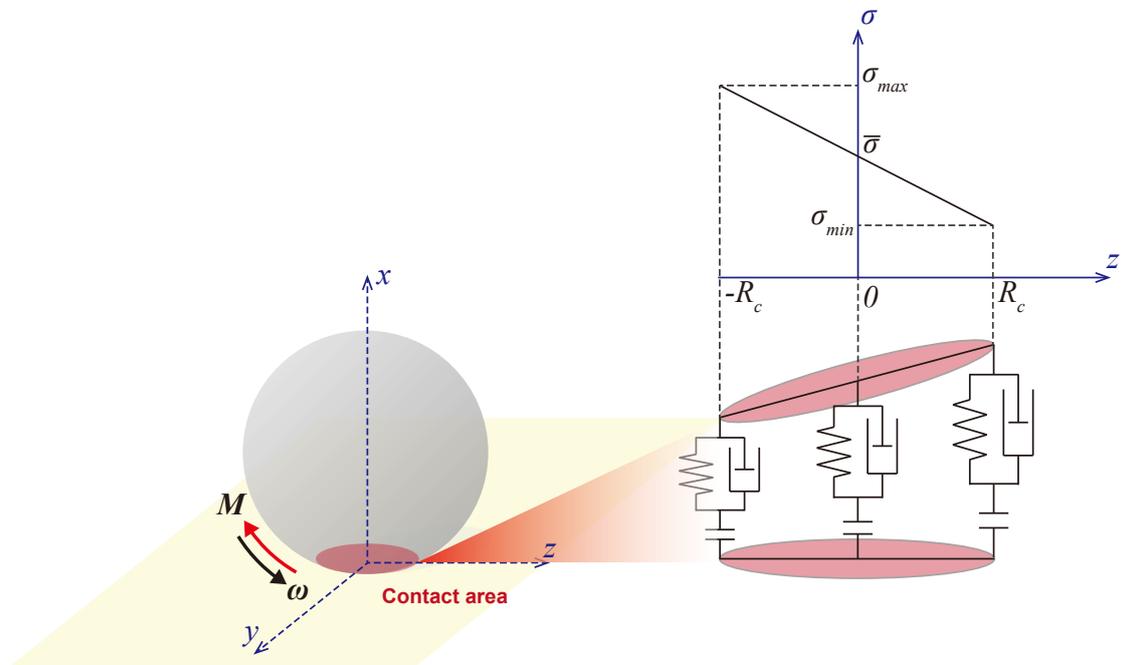

**Fig. 2 Force balance of a particle on an inclined plane.**

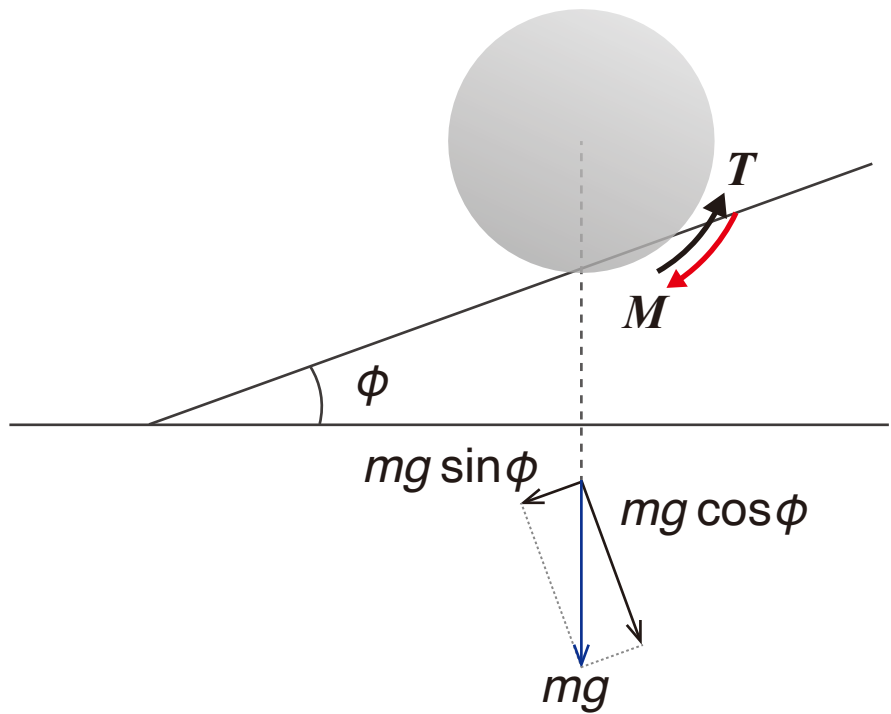

**Fig. 3 Overview of the coarse-grained model incorporating rolling resistance model.**

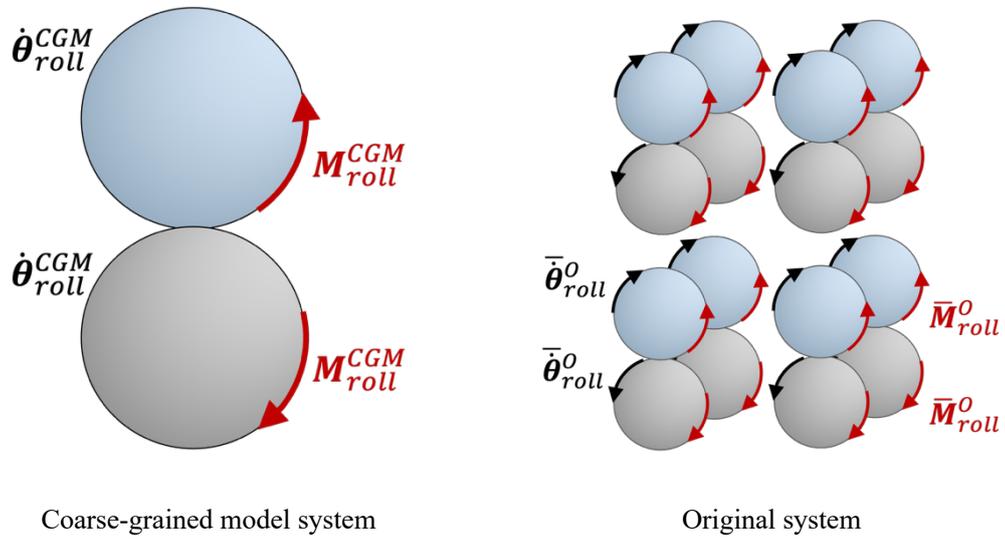

Coarse-grained model system     Original system

Fig. 4 Initial configuration of granular assembly for stability test of the proposed rolling resistance model.

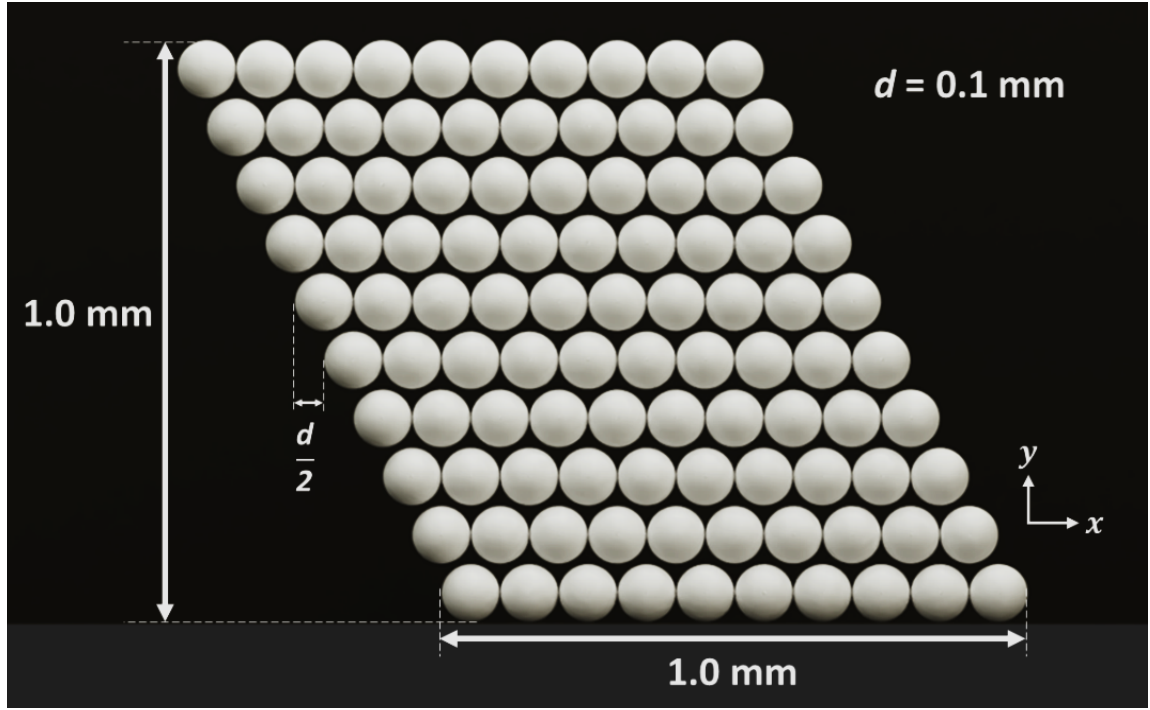

**Fig. 5 Snapshots of particle assembly at 5.0 and 8.0 s ((a) Case 1-1: the proposed rolling resistance model, (b) Case 1-2: the existing model).**

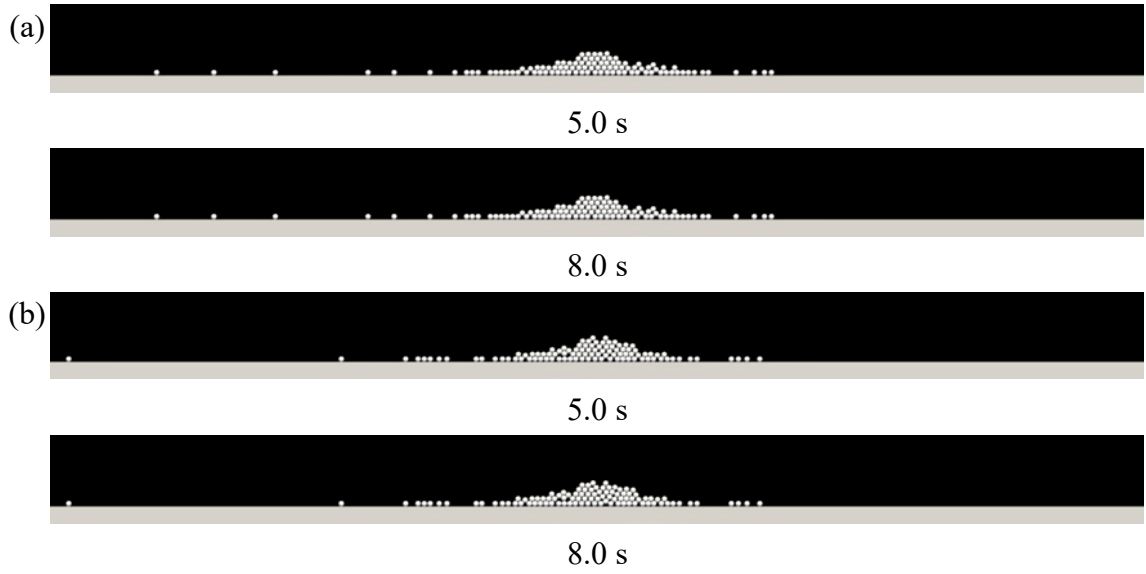

**Fig. 6 Absolute angular velocity distribution of particle assembly at 5.0 and 8.0 s ((a) Case 1-1: the proposed rolling resistance model, (b) Case 1-2: the existing model).**

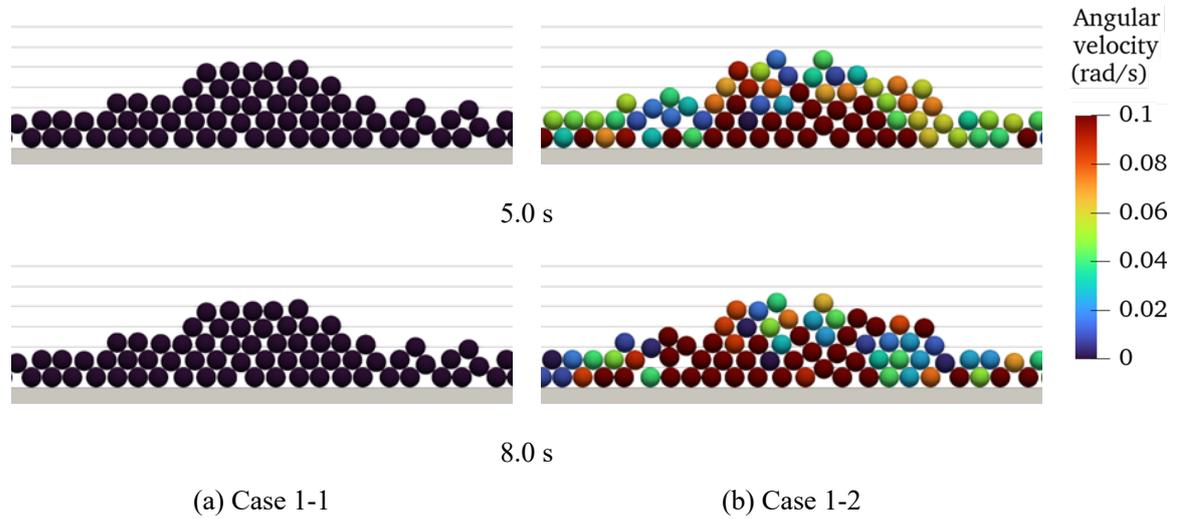

Fig. 7 Transient change in the average absolute angular velocity of particles from 4.0 to 8.0 seconds in Cases 1-1 (the proposed model) and 1-2 (the existing model).

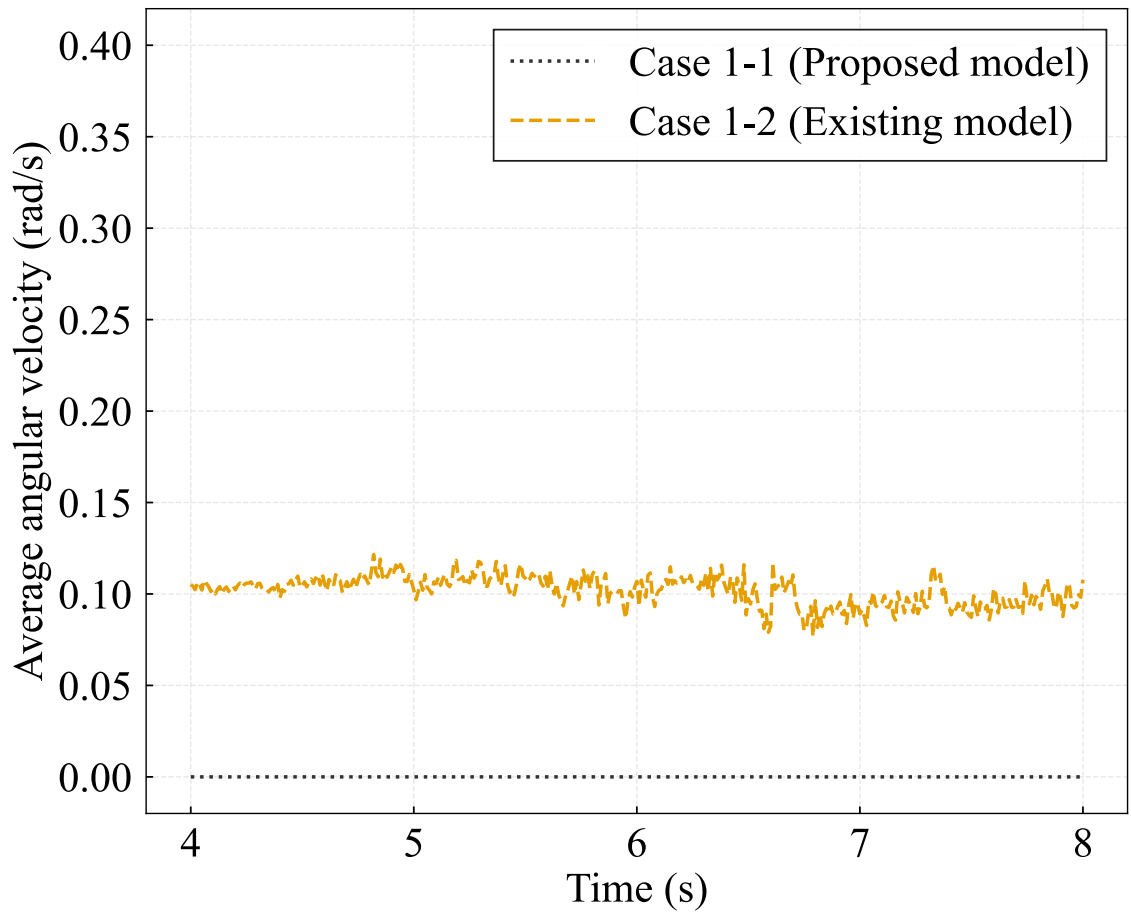

**Fig. 8 Snapshots of particle assembly at 6.0 and 8.0 s ((a) Case 1-3: the proposed rolling friction model, (b) Case 1-4: the existing model).**

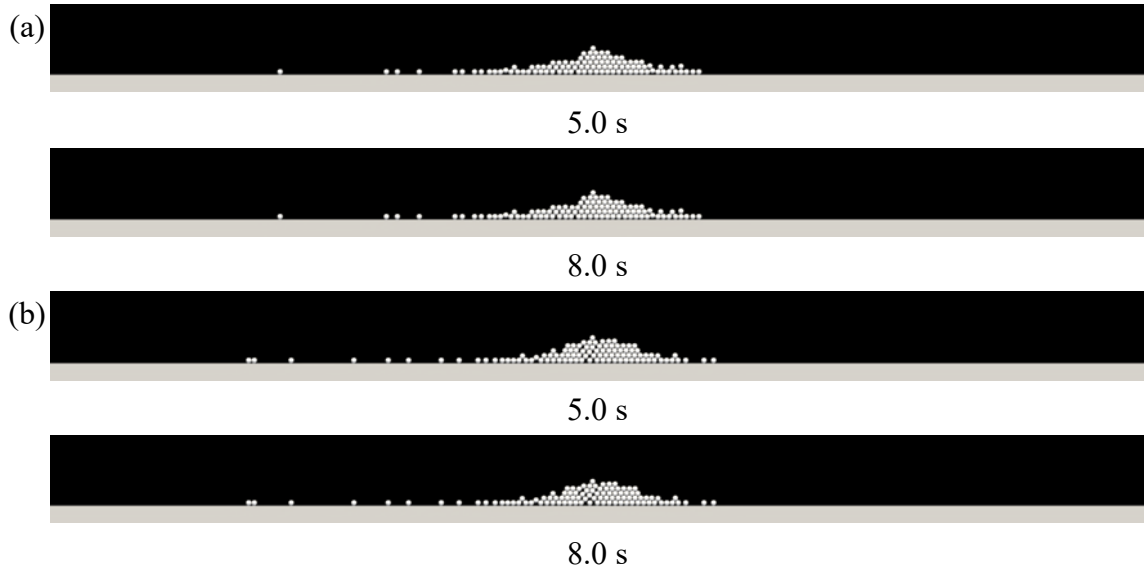

**Fig. 9 Absolute angular velocity distribution of particle assembly at 5.0 and 8.0 s ((a) Case 1-3: the proposed rolling resistance model, (b) Case 1-4: the existing model).**

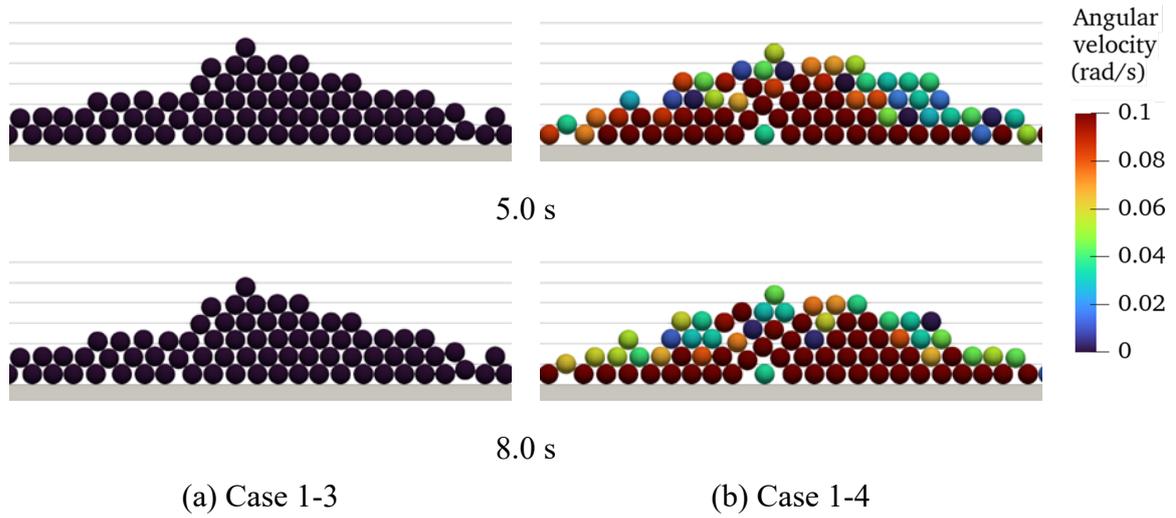

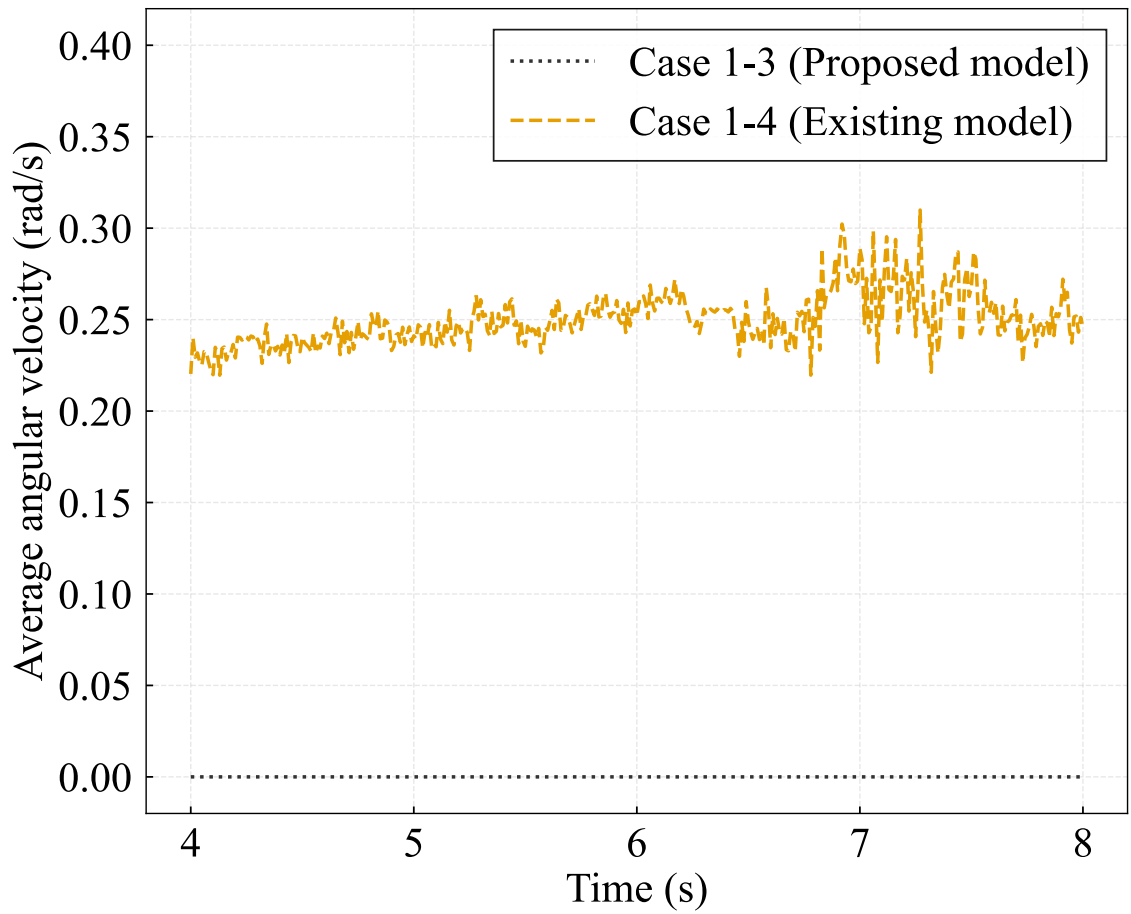

Fig. 10 Transient change in the average absolute angular velocity of particles from 5.0 to 8.0 seconds in Cases 1-3 (the proposed model) and 1-4 (the existing model).

**Fig. 11 Schematic of an incinerator and the initial location of the particles.**

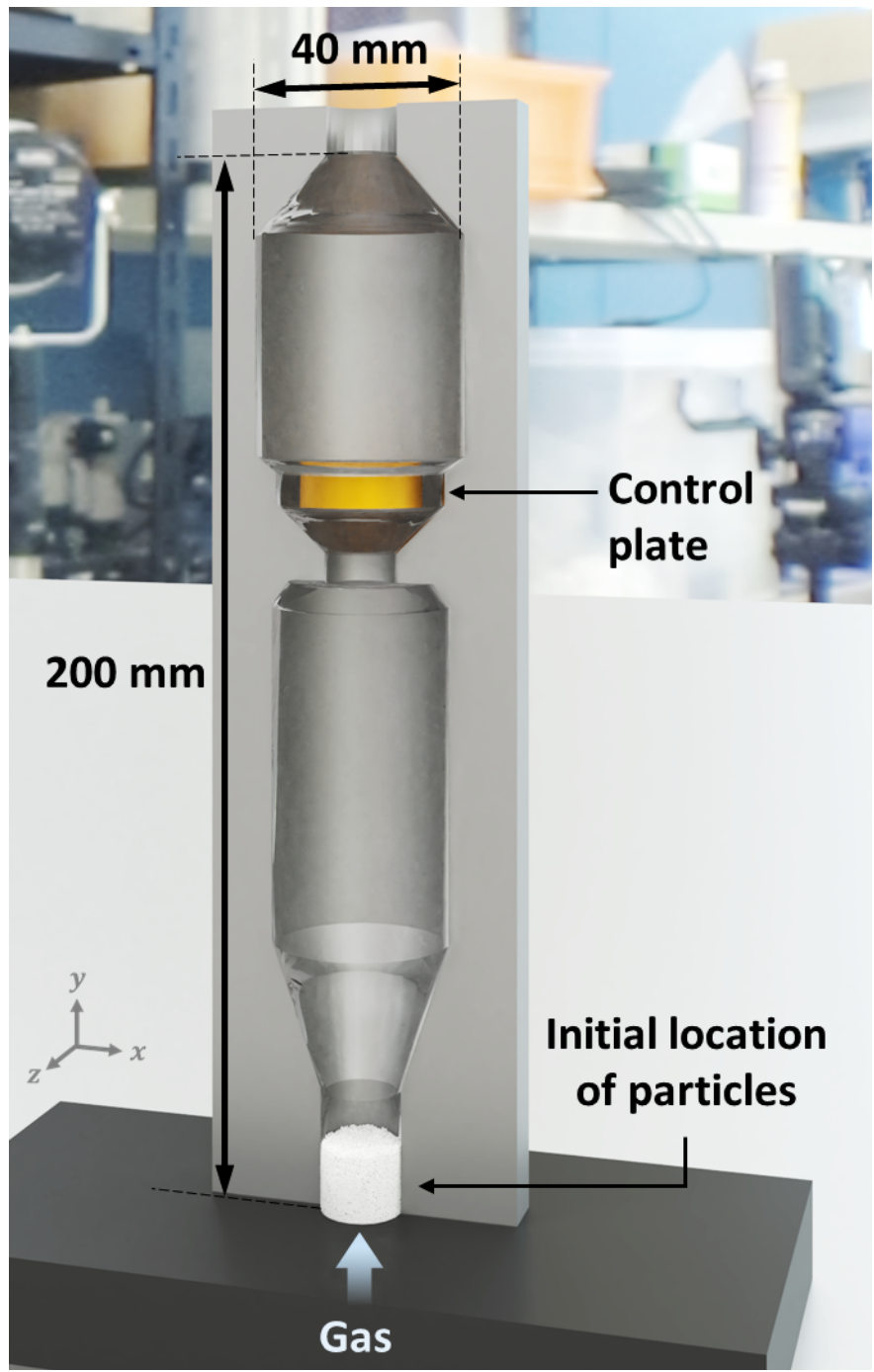

**Fig. 12 Typical snapshots of particle behavior in the incinerator (Case 2).**

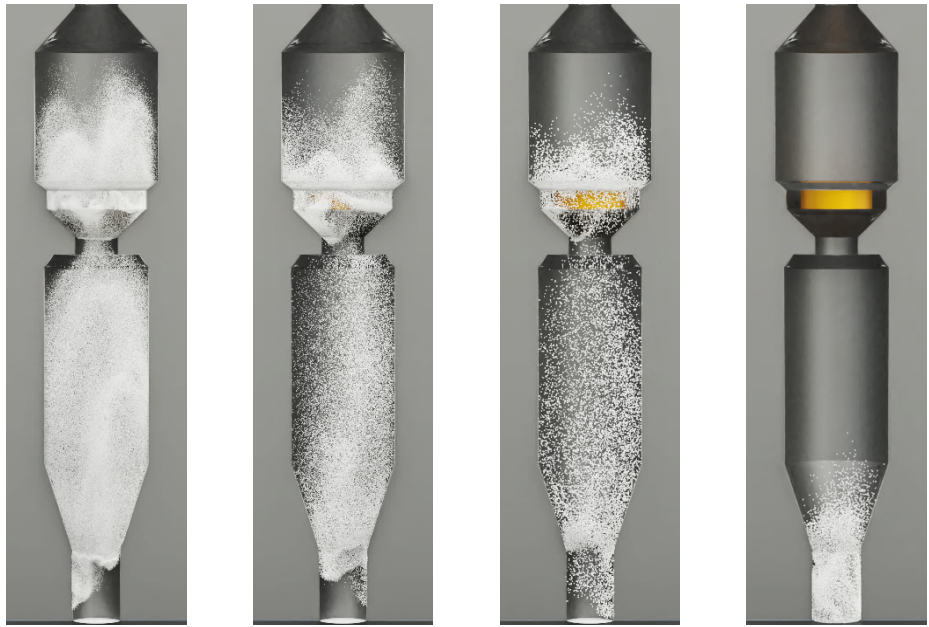

(a) Early stage

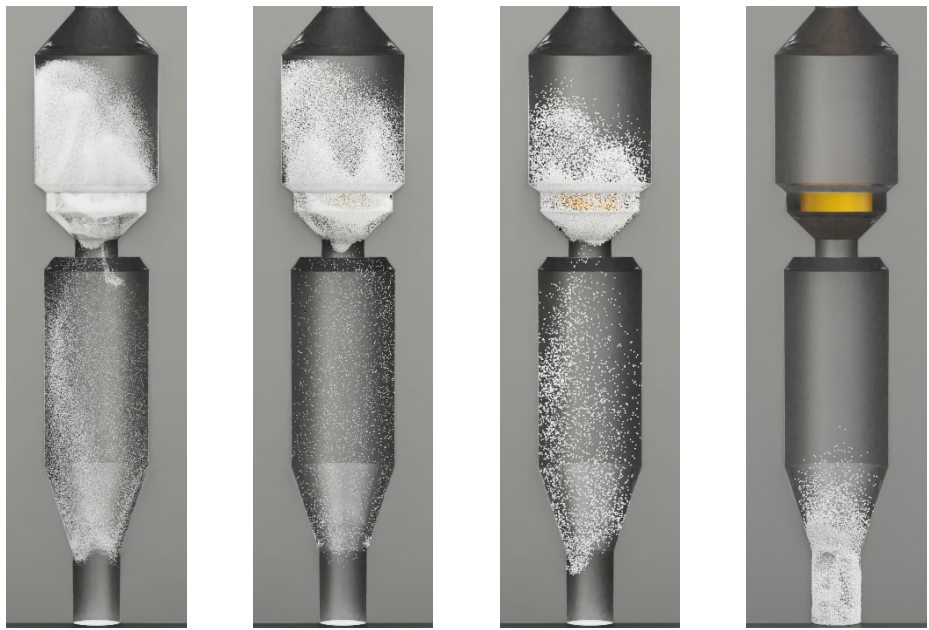

(b) Later stage

Case 2-1    Case 2-2    Case 2-3    Case 2-4

**Fig. 13 Typical velocity distribution of particle and gas in the incinerator (Case 2).**

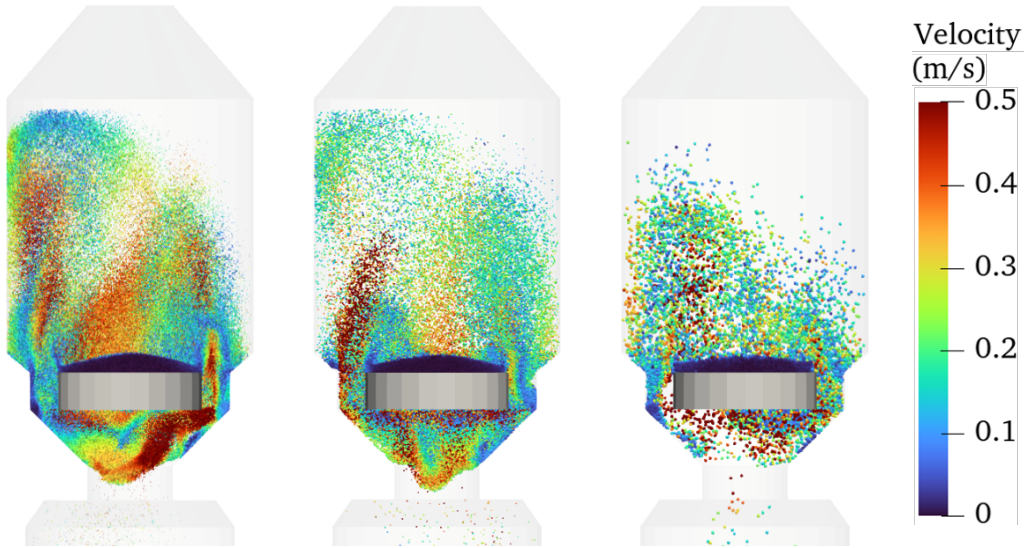

(a) Half-cut view particle velocity distribution

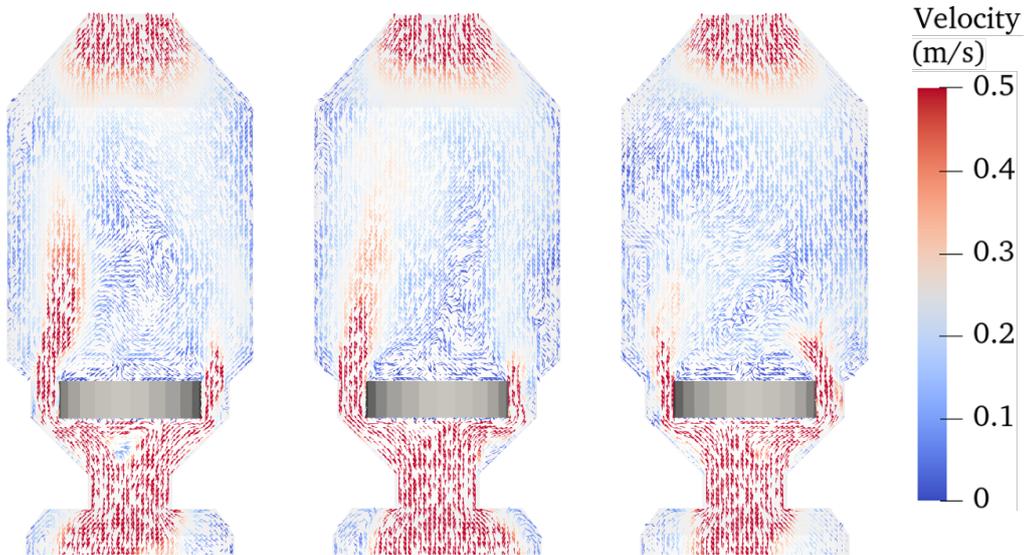

(b) Air-flow vector at the central slice

Case 2-1          Case 2-2          Case 2-3

**Fig. 14 Transient change of the total kinetic energy in the incinerator (Case 2).**

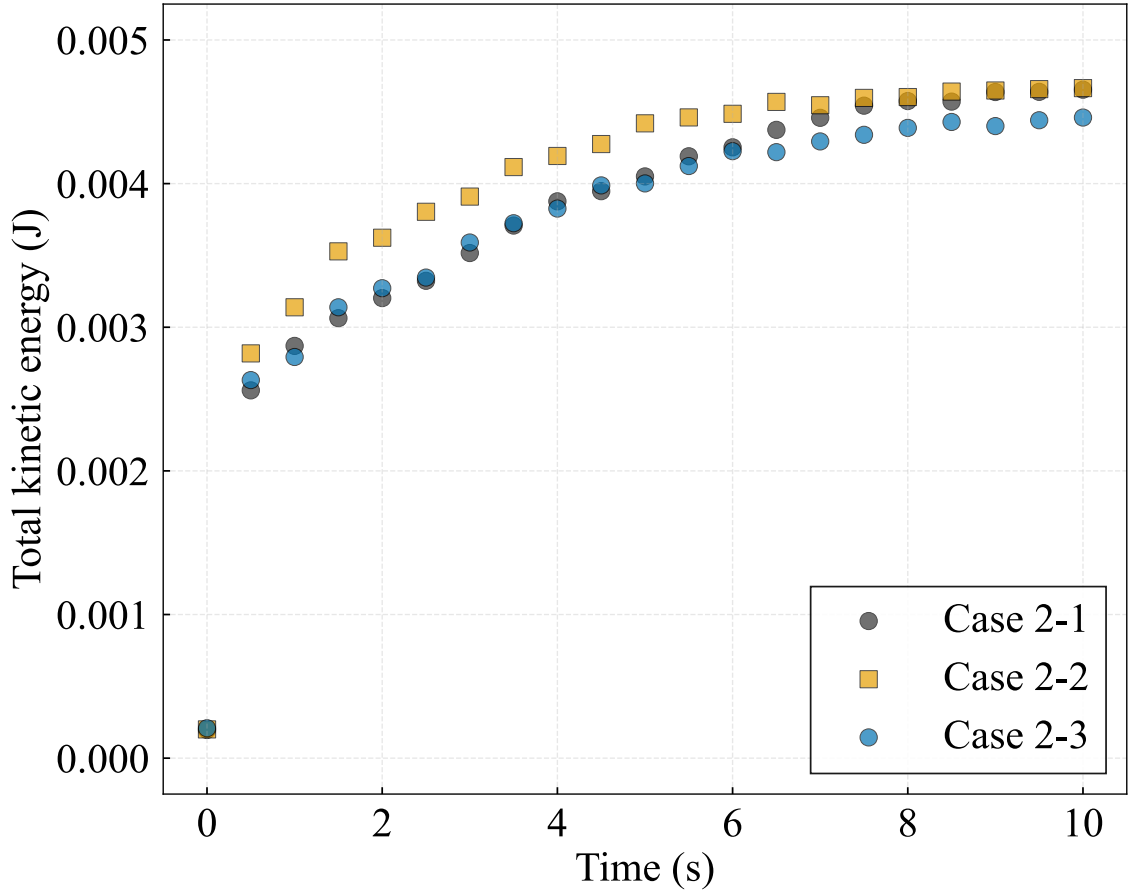

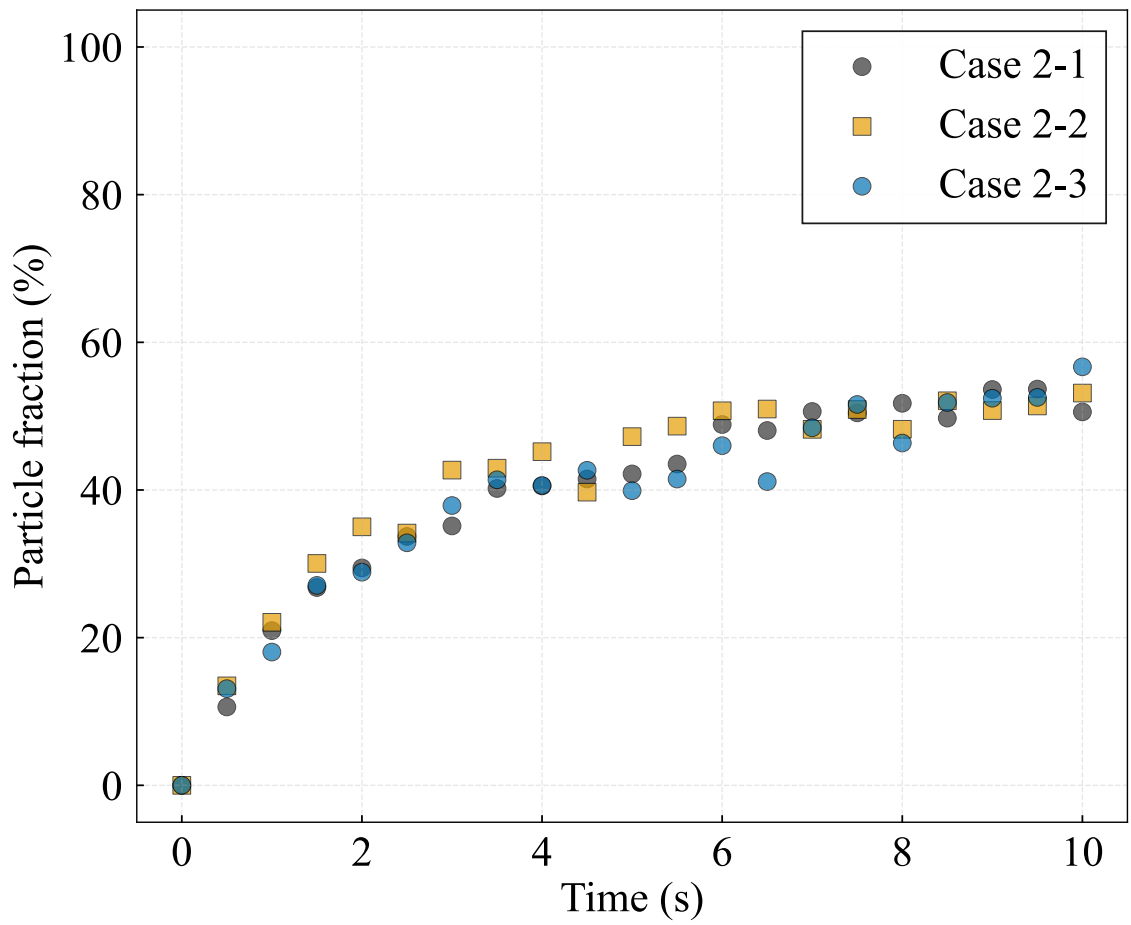

Fig. 15 Transient change in the particle fraction in the region above the control plate in the incinerator (Case 2).

**Fig. 16 Pressure drop in the incinerator (Case 2).**

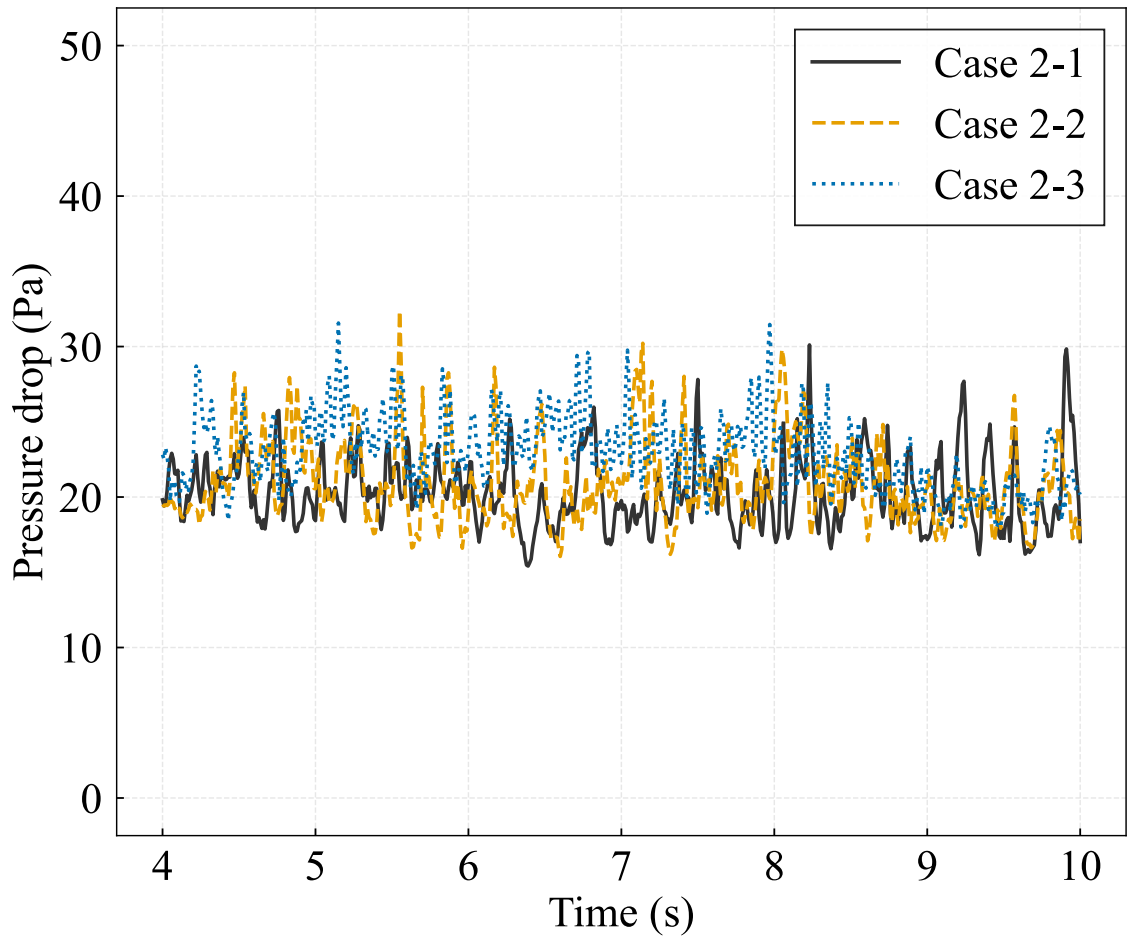

**Fig. 17 Snapshot of particles on the control plate at 10.0 s (Case 2).**

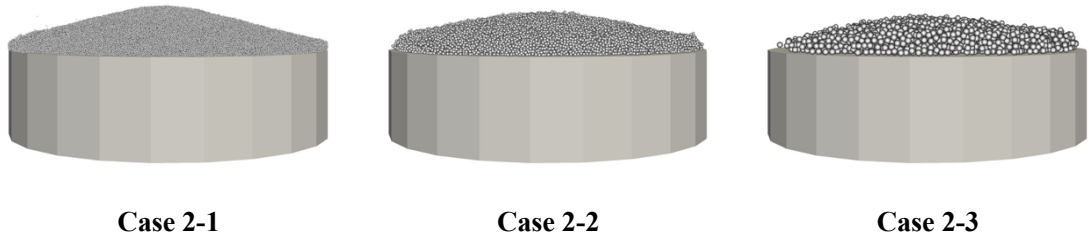

       **Case 2-1**                 **Case 2-2**                 **Case 2-3**

Fig. 18 Typical snapshots of particle behavior in the incinerator (Case 3).

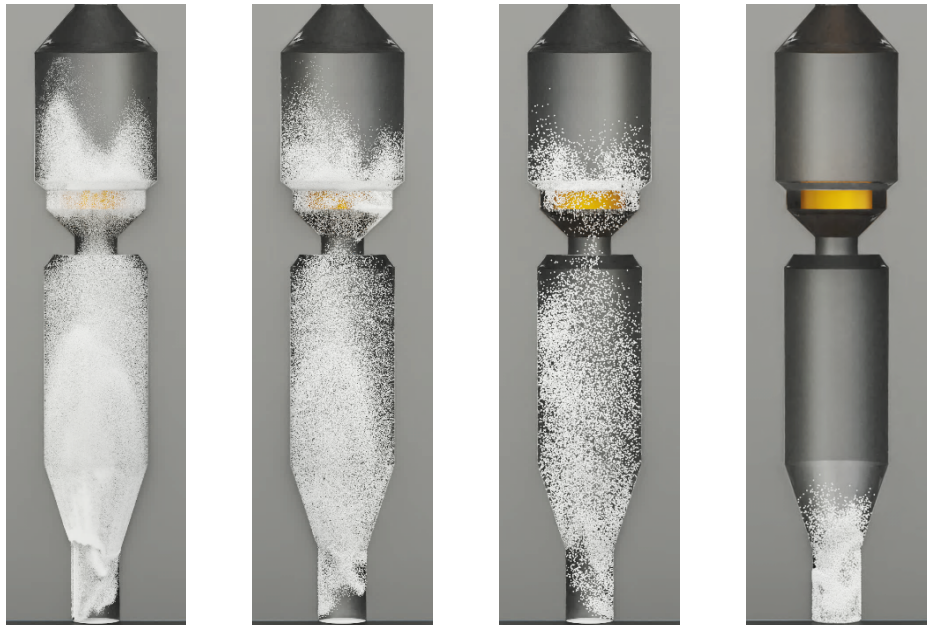

Early stage

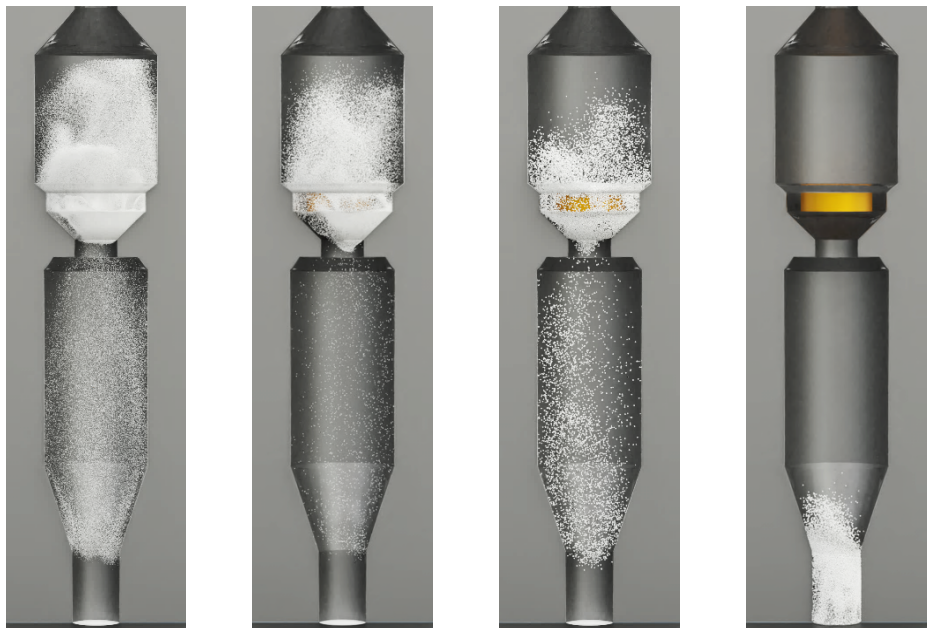

Later stage

Case 3-1    Case 3-2    Case 3-3    Case 3-4

**Fig. 19 Typical velocity distribution of particle and gas in the incinerator (Case 3).**

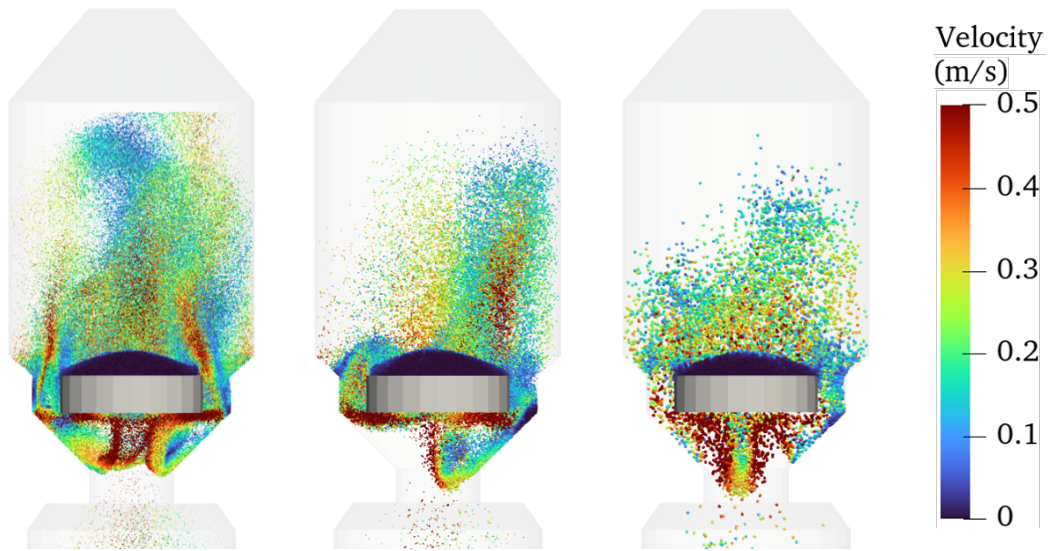

(a) Half-cut view particle velocity distribution

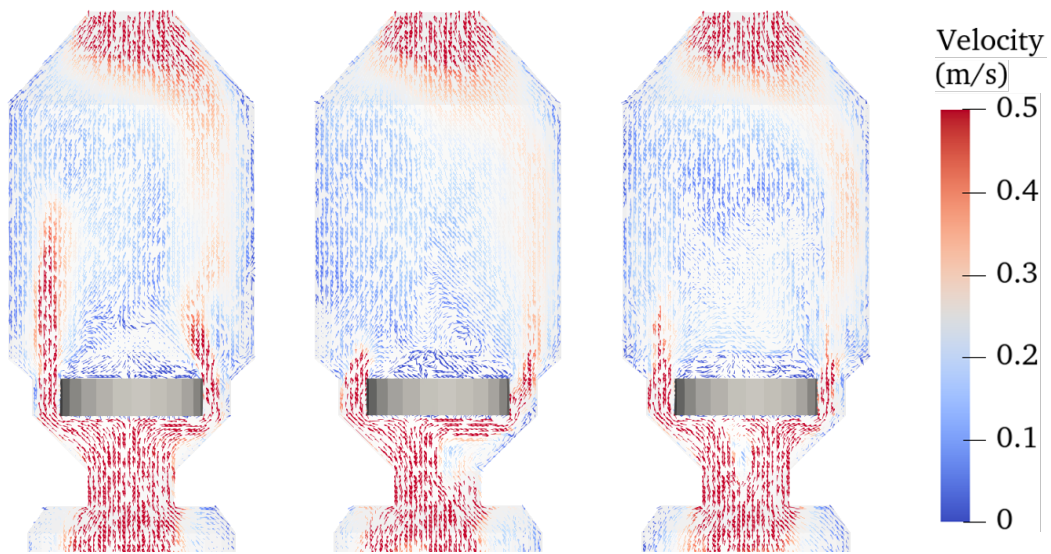

(b) Air-flow vector at the central slice

Case 3-1　　　　Case 3-2　　　　Case 3-3

**Fig. 20 Transient change of the total kinetic energy in the incinerator (Case 3).**

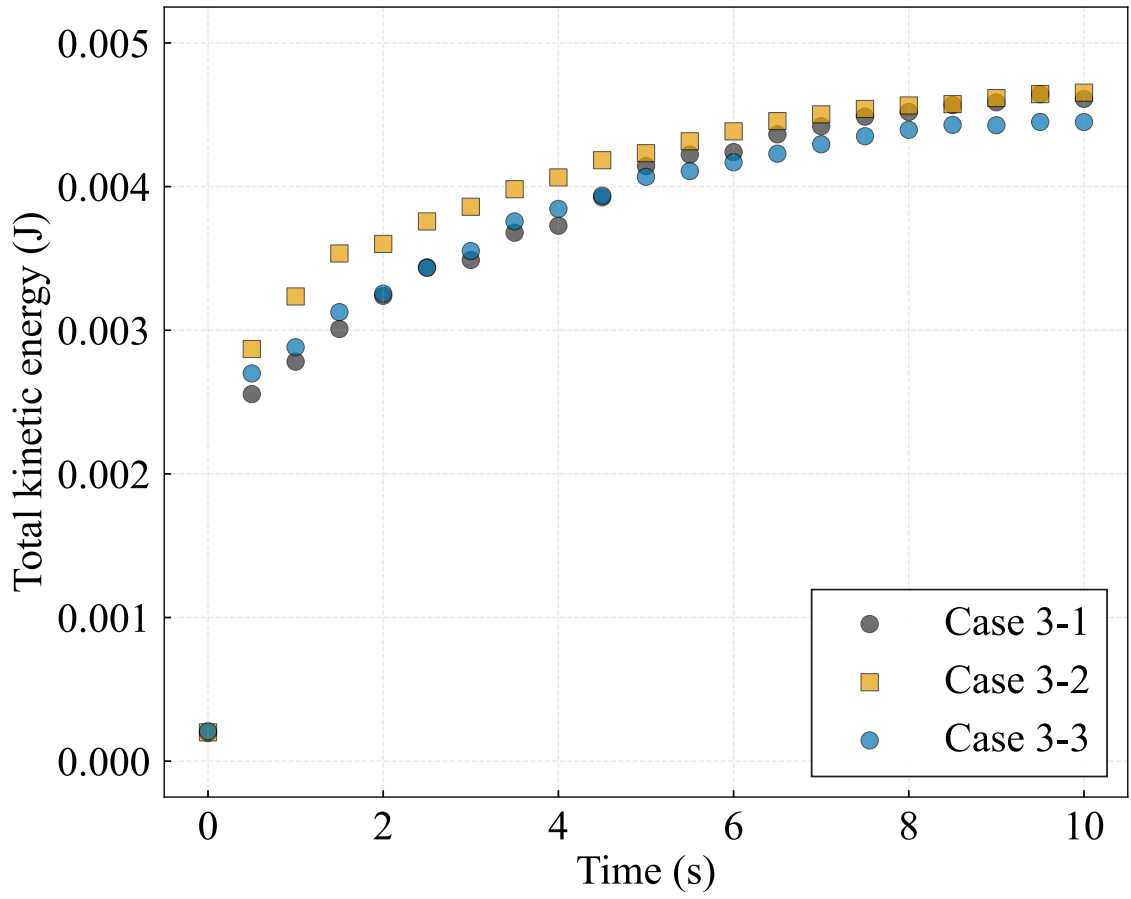

**Fig. 21** Transient change in the particle fraction in the region above the control plate in the incinerator (Case 3).

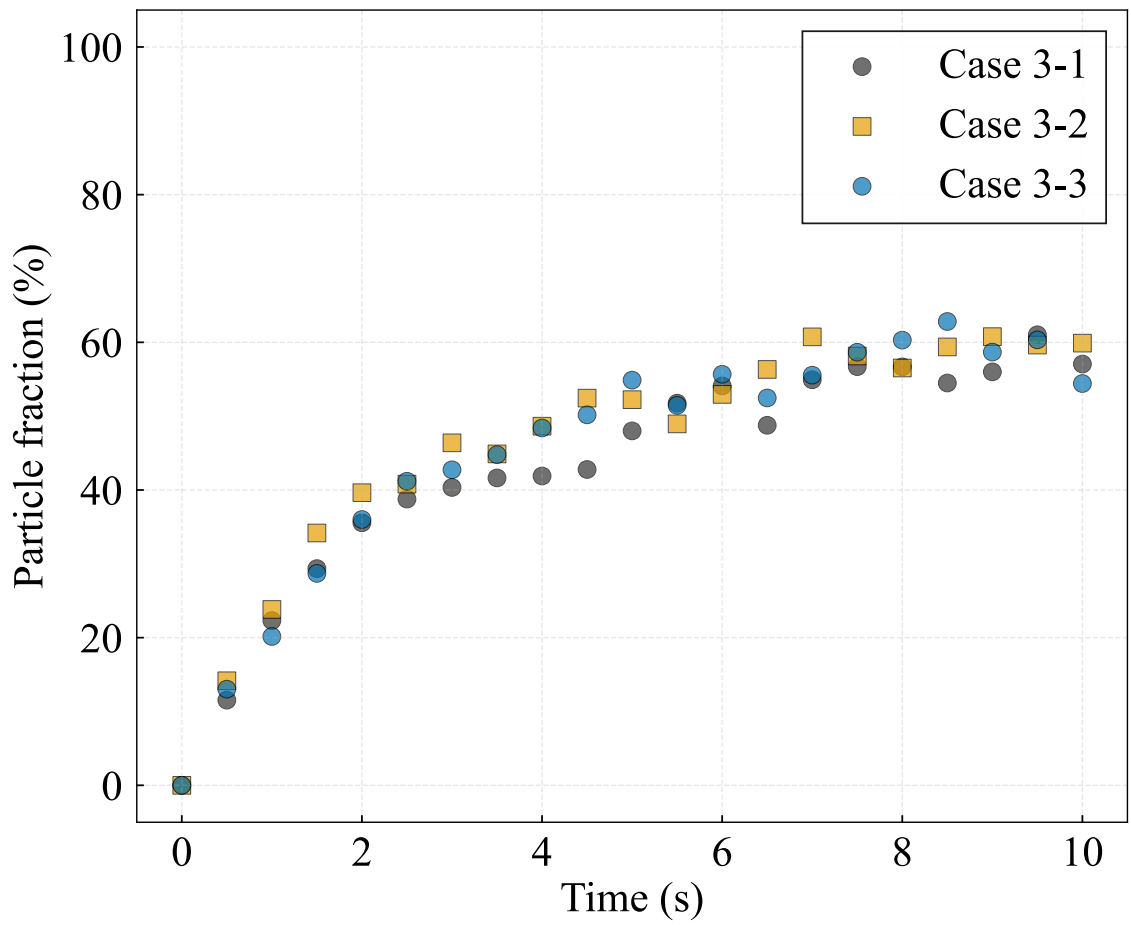

**Fig. 22 Pressure drop in the incinerator (Case 3).**

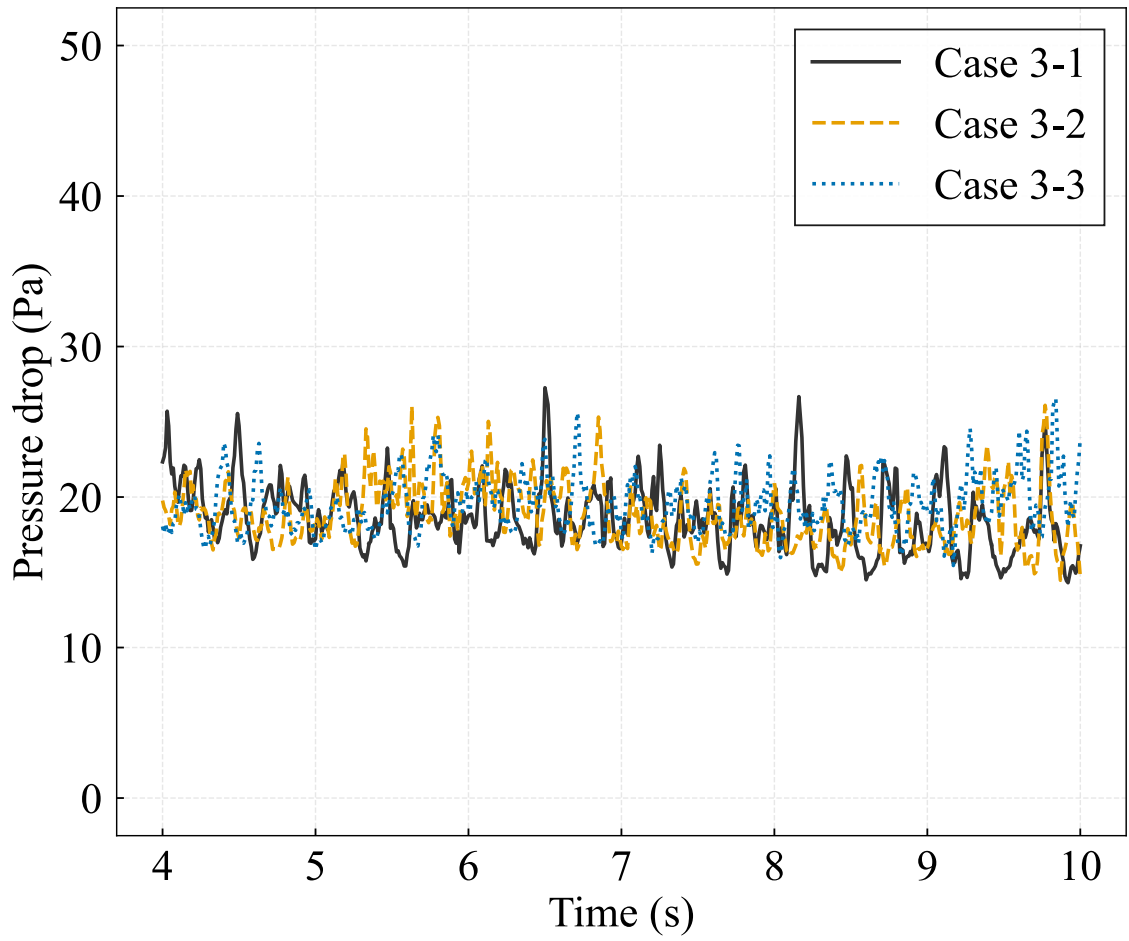

**Fig. 23 Snapshot of particles on the control plate during the quasi-steady state (Case 3).**

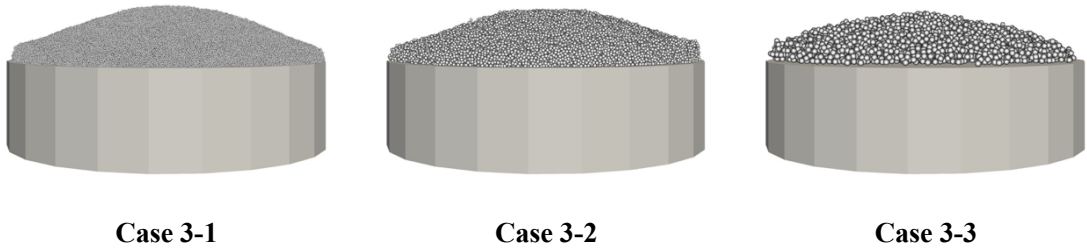

        **Case 3-1**                **Case 3-2**              **Case 3-3**

**Table 1 Physical properties in the stability tests.**

| Particle density | kg/m$^3$ | 2,500 |
|---|---|---|
| Spring constant | N/m | 100 |
| Coefficient of restitution | – | 0.9 |
| Coefficient of friction | – | 0.3 |

**Table 2 Physical properties in the validation tests.**

| | | |
|---|---|---|
| *Gas phase* | | |
| Density | kg/m$^3$ | 1.0 |
| Viscosity | Pa・s | $1.8\times10^{-5}$ |
| *Solid phase* | | |
| Particle density | kg/m$^3$ | 2,500 |
| Spring constant | N/m | 100 |
| Coefficient of restitution | – | 0.9 |
| Coefficient of friction | – | 0.3 |

**Table 3 Calculation conditions in the validation tests.**

|  |  | Case 2-1 | Case 2-2 | Case 2-3 | Case 2-4 | Case 3-1 | Case 3-2 | Case 3-3 | Case 3-4 |
|---|---|---|---|---|---|---|---|---|---|
| Original particle size | μm | 100 | 100 | 100 | 400 | 100 | 100 | 100 | 400 |
| Calculated particle size | μm | 100 | 200 | 400 | 400 | 100 | 200 | 400 | 400 |
| Coarse-grain ratio | – | 1.0 | 2.0 | 4.0 | 1.0 | 1.0 | 2.0 | 4.0 | 1.0 |
| Number of particles | – | 2,560,000 | 320,000 | 40,000 | 40,000 | 2,560,000 | 320,000 | 40,000 | 40,000 |
| Critical rolling angle | rad | – | | | | 0.1 | | | |

**Table 4 Simulation results of angle of repose measurements.**

|  | Angle of repose (°) |
|---|---|
| Without rolling friction |  |
| Case 2-1 (Coarse-grain ratio: 1.0/ Original) | 13.62 ± 0.99 |
| Case 2-2 (Coarse-grain ratio: 2.0) | 11.70 ± 1.51 |
| Case 2-3 (Coarse-grain ratio: 4.0) | 7.13 ± 1.95 |
| With rolling friction |  |
| Case 3-1 (Coarse-grain ratio: 1.0/ Original) | 20.02 ± 1.72 |
| Case 3-2 (Coarse-grain ratio: 2.0) | 18.42 ± 1.45 |
| Case 3-3 (Coarse-grain ratio: 4.0) | 18.40 ± 4.50 |